\documentclass[11pt, a4paper]{article}
\pdfoutput=1 

\usepackage[height=8.85in,width=6.6in]{geometry}

\usepackage{graphicx,rotating}     
\usepackage[bookmarksopen,colorlinks=true,linkcolor=light_blue,
citecolor=light_pink,urlcolor=dark_red,linktoc=all]{hyperref}

\usepackage{amsmath}
\usepackage{mathtools}
\usepackage{amsfonts}
\usepackage{amssymb}
\usepackage{slashed}
\usepackage{braket}
\usepackage{bm}
\usepackage{cite}
\usepackage{comment}
\usepackage{physics}
\usepackage{xcolor}
\usepackage[utf8]{inputenc}
\usepackage[footnotesize]{caption}
\usepackage{bbold}

\usepackage{fourier}

\newcommand{\Ham}{\mathcal{H}}

\numberwithin{equation}{section}

\usepackage{multicol}
\definecolor{dark_red}{rgb}{0.7, 0., 0.}
\definecolor{light_pink}{rgb}{1,0.4,0.4}
\definecolor{light_blue}{rgb}{0.284602,0.317763,0.963947}
\definecolor{cred}{RGB}{180,50,40} 
\definecolor{darkgreen}{RGB}{0, 100, 0}
\definecolor{desy_blue}{HTML}{009EE2}
\definecolor{desy_orange}{HTML}{FD8800}
\definecolor{forestgreen}{HTML}{228B22}
\definecolor{ochre}{HTML}{CCAA2B}

\begin{document}

\hypersetup{pageanchor=false}
\begin{titlepage}

\begin{center}

\hfill CERN-TH-2022-075 \\
\hfill KEK-TH-2422

\vskip 1.in

{\Huge \bfseries 
Charge transfer between \\
rotating complex scalar fields\\
}
\vskip .8in

{\Large Valerie Domcke$^{a,b}$, Keisuke Harigaya$^{a,c}$, Kyohei Mukaida$^{d,e}$}

\vskip .3in
\begin{tabular}{ll}
$^a$& \!\!\!\!\!\emph{Theoretical Physics Department, CERN, 1211 Geneva 23, Switzerland}\\
$^b$& \!\!\!\!\!\emph{Laboratory for Particle Physics and Cosmology, Institute of Physics, }\\[-.3em]
& \!\!\!\!\!\emph{School of Basic Sciences, EPFL, 1015 Lausanne, Switzerland}\\
$^c$ & \!\!\!\!\!\emph{Kavli IPMU (WPI), UTIAS, The University of Tokyo, Kashiwa, Chiba 277-8583, Japan}\\
$^d$& \!\!\!\!\!\emph{Theory Center, IPNS, KEK, 1-1 Oho, Tsukuba, Ibaraki 305-0801, Japan}\\
$^e$& \!\!\!\!\!\emph{Graduate University for Advanced Studies (Sokendai), }\\[-.3em]
& \!\!\!\!\!\emph{1-1 Oho, Tsukuba, Ibaraki 305-0801, Japan}
\end{tabular}

\end{center}
\vskip .6in

\begin{abstract}
\noindent We consider the transfer of a $U(1)$ charge density between Bose-Einstein condensates of complex scalar fields coupled to a thermal bath, focusing on the case of a homogeneous Affleck-Dine field transmitting the charge stored in its angular motion to an axion field. We demonstrate that in the absence of additional symmetries this charge transfer, aided by cosmic expansion as well as the  thermal effective  potential of the Affleck-Dine field, can be very efficient. The charge redistribution between the scalar fields becomes possible if the interactions with the thermal bath break the original $U(1) \times U(1)$ symmetry down to a single $U(1)$ symmetry; the charge distribution between the two fields is then determined by minimizing the free energy. We discuss implications for cosmological setups involving complex scalars, with applications to axion dark matter, baryogenesis, kination domination, and gravitational wave production.
\end{abstract}

\end{titlepage}

\tableofcontents
\thispagestyle{empty}
\renewcommand{\thepage}{\arabic{page}}
\renewcommand{\thefootnote}{$\natural$\arabic{footnote}}
\setcounter{footnote}{0}
\newpage
\hypersetup{pageanchor=true}

\newpage
\section{Introduction}
\label{sec:intro}

Dynamical complex scalar fields may play an important role in the early Universe. Here we consider (composite) complex scalar fields subject to an approximate $U(1)$ symmetry. For a homogeneous field configuration, the corresponding approximately conserved charge corresponds to a rotation of the complex scalar in field space, i.e., a non-vanishing velocity for the angular degree of freedom.
The angular motion may be a source of dark matter generation~\cite{Co:2019jts}.
Moreover, this motion spontaneously breaks $CP$ and can thus contribute to the generation of a matter antimatter asymmetry~\cite{Sakharov:1967dj}. Concrete realizations of this are the Affleck-Dine (AD) mechanism~\cite{Affleck:1984fy}, spontaneous baryogenesis~\cite{Cohen:1987vi,Cohen:1988kt}, or axiogenesis~\cite{Co:2019wyp}.

In AD baryogenesis, the role of an rotating complex AD field is naturally played by a condensate of scalar particles aligned along a flat direction of the scalar potential of the minimal supersymmetric standard model (MSSM). See \cite{Gherghetta:1995dv} for the list of the MSSM flat directions and \cite{Enqvist:2003gh} for a review on their dynamics. The rotation is induced by higher-dimensional baryon or lepton number-violating operators and the charge associated with this rotation is transferred to thermal bath through the decay of the AD field into Standard Model (SM) fermions. 
In the more recently proposed axiogenesis scenario, the rotating complex field is instead a fundamental scalar, namely the complex Peccei-Quinn (PQ) field introduced to explain the absence of any observed $CP$ violation in QCD~\cite{Peccei:1977hh,Peccei:1977ur}. The angular degree of freedom is referred to as the axion~\cite{Weinberg:1977ma,Wilczek:1977pj}, which simultaneously provides a promising dark matter candidate~\cite{Preskill:1982cy,Abbott:1982af,Dine:1982ah}. The charge transfer between the rotating axion and the SM thermal bath occurs via sphaleron interactions. The kinetic energy of the axion field arising from the rotation of the PQ field is transferred into an axion dark matter density through the kinetic misalignment mechanism~\cite{Co:2019jts}.

In this paper, we study the charge transfer between Bose-Einstein condensates (BECs) of different complex scalar fields in the presence of a thermal bath. For two or more scalar fields coupled through efficient interactions with a thermal bath, the $U(1)$ charge will be redistributed among the fields and the thermal bath to minimize the free energy. If after coupling to the thermal bath, the system has more than one relevant conserved charge, this additional symmetry implies that a charge transfer between the scalar fields would be accompanied by a large chemical potential in the thermal bath so that the free energy condition disfavours a charge transfer. On the other hand, in the absence of additional symmetries we find that the charge transfer can be very efficient, and must be taken into account in cosmologies involving rotating complex scalar fields.
Our work is related to earlier works that discuss the generation of a non-zero velocity of an axion field by asymmetries of other particles or fields, such as a quark chiral asymmetry~\cite{McLerran:1990de}, helical magnetic fields~\cite{Kobayashi:2020bxq}, and a baryon asymmetry~\cite{Alonso:2020com}.

Concretely, we study a system of two complex scalar fields coupled to a thermal bath via sphaleron, Yukawa, and gauge interactions. We consider one of the fields, the `AD field', to have a large initial charge encoded in an angular motion. The second field, the `PQ field', is initially at rest at the minimum of its zero-temperature Mexican-hat scalar potential with the PQ symmetry spontaneously broken. Minimizing the free energy of this system, we demonstrate that in the ground state of the system, the charge has largely been transferred to the axion field (i.e., the angular component of the PQ field) unless additional conserved symmetries (such as a chiral symmetry for the fermions of the thermal bath) require that the transfer involve a large fermion asymmetry. This charge transfer occurs once the transfer rates governing this process are efficient compared to the Hubble expansion rate and once the charge associated with the AD field rotation has been red-shifted close to a critical value. At earlier times, the system is trapped in a state with the charge largely stored in the rotation of the AD field. For sufficiently large initial values for the AD field, a phase of AD field domination with a subsequent kination era by the axion field is possible.

Thermal contributions to the effective potential governing the dynamics of the AD field and the axion can further facilitate the charge transfer. We compute the relevant thermodynamic quantities in the presence of rotating scalar fields to obtain the one-loop thermal effective potential of the coupled system of the AD field, axion, and thermal bath. If the thermal bath contains particles which obtain zero-temperature masses from a finite AD field value (e.g., due to a Yukawa coupling or due to the spontaneous breaking of a gauge symmetry by the AD field), the resulting contribution to the effective potential prefers a vanishing AD field value. When the total charge is sufficiently small (but is still above the critical value), this creates an absolute minimum at the origin, while the state with a large AD field value remains a local minimum. This is analogous to the shape of the potential that appears in thermal inflation models~\cite{Yamamoto:1985rd,Lyth:1995ka}. As Hubble expansion dilutes the charge of the coupled system, the latter vacuum is eventually destabilized and all the remaining charge is transferred to the axion field. This process may occur via nucleation of bubbles of true vacuum in which the axion field is rotating~\cite{Coleman:1977py,Callan:1977pt,Linde:1981zj} or through a more gradual phase mixing driven by sub-critical bubble formation~\cite{Gleiser:1991rf,Dine:1992wr,Shiromizu:1995jv,Hiramatsu:2014uta}. The former will generate  gravitational waves.
In both cases, fluctuations of the  axion field are created, which contribute to axion dark matter.
Also, the formation of Q-balls~\cite{Coleman:1985ki,Laine:1998rg} is possible, and the fluctuations associated with the Q-ball formation and decay may lead to an additional source of axion dark matter.

The remainder of this paper is organized as follows. In Sec.~\ref{sec:zero}, we discuss charge transfer neglecting the thermal corrections to the potential of the AD field and clarify the conditions for the transfer to occur by computing the effective potential for a given finite charge density. We demonstrate how additional conserved charges can prevent the charge transfer.  In Sec.~\ref{sec:tunneling}, we include the thermal correction to the AD field potential and show how the shape of the potential is modified. Phenomenological implications of the charge transfer are discussed in Sec.~\ref{sec:pheno}. Finally, we give a summary and discussion in Sec.~\ref{sec:conclusion}. Technical details can be found in the appendices. App.~\ref{app:Boltzmann} provides numerical solution for the Boltzmann equations of Sec.~\ref{sec:zero}. A derivation of the effective potential of the Affleck Dine field, key to the discussion in Sec.~\ref{sec:tunneling}, is given in App.~\ref{app:Veff}.


\section{Charge transfer in the zero-temperature potential}
\label{sec:zero}

In this section, we discuss the charge transfer between a homogeneous complex scalar field $\phi$ and an axion field $a = f \theta_a$
and illustrate the conditions for this charge transfer to occur efficiently. Throughout this paper, we assume that the radial direction of the complex PQ symmetry breaking field $P$ is fixed at its VEV$\sim f$, so that we can integrate it out and use the effective field theory of the axion field without the radial direction,
\begin{align}
  P = \frac{1}{\sqrt{2}} f e^{i \theta_a} \,.
\end{align}
In particular, we discuss the dynamics of the axion at $T \ll f, |\phi|$, where both complex scalar fields can be described as coherently rotating in field space.
This does however not exclude a restoration of the PQ symmetry at higher temperatures.
The era with $|\phi| <T$ is discussed in detail in Sec.~\ref{sec:tunneling}.

We assume that the complex scalar field $\phi$ initially rotates with a large radius. The rotation can be initiated by the Affleck-Dine (AD) mechanism~\cite{Affleck:1984fy}, and we call the rotating scalar field $\phi$ the AD field. We refer to the radial and angular directions of $\phi$ as $r$ and $\theta_\phi$, respectively,
\begin{align}
    \phi = \frac{1}{\sqrt{2}}r e^{i \theta_\phi}.
\end{align}
To be concrete, we assume that the zero temperature potential of the AD field is nearly quadratic, $m^2|\phi|^2$, which is indeed the case when the AD field is a flat direction in the MSSM. See~\cite{Gherghetta:1995dv} for the list of MSSM flat directions.
Throughout this paper we will assume $m \ll f$, motivated e.g., by the QCD axion with an axion decay constant constrained to $f > 10^8$~GeV and low scale supersymmetry with a supersymmetry breaking scale of $m \sim$~TeV. 
In realistic cases, the zero temperature potential is not exactly quadratic. Indeed, the MSSM flat directions receive logarithmic corrections by renormalization group effects,
\begin{align}
    V(\phi) = m^2 |\phi|^2 \qty( 1 + K \, {\rm ln} \frac{\phi^2}{M^2}),
\end{align}
where $K$ is a constant of $O(10^{-3}-10^{-1})$ and $M$ is a mass scale. We assume $K>0$ so that the rotation is stable against fragmentation into Q-balls~\cite{Copeland:2009as}.
The potential of the AD field moreover receives thermal corrections, which we will discuss in Sec.~\ref{sec:tunneling}.

We introduce interactions that can transfer the charge of the AD field to the axion field, inducing  rotations of the axion field with almost all of the charges eventually transferred to the axion.
For simplicity, we will take the axion field to be at rest initially, $\dot \theta_a = 0$.
We discuss several models to illustrate the conditions for the efficient transfer to occur. In particular, Sec.~\ref{sec:gauge} computes the charge transfer rate in a situation where the charge transfer can occur and Sec.~\ref{sec:fermion} shows how additional (approximate) symmetries can prevent the charge transfer.

\subsection{Charge transfer through an anomalous coupling to a gauge field}
\label{sec:gauge}
\paragraph{Setup and conserved charge.} We first consider a simple example that will illustrate the condition on the free energy for the transfer to occur.
We introduce a non-Abelian gauge field $G$ that couples to the AD and axion fields through a quantum anomaly,
\begin{align}
\label{eq:model1}
    \left(\theta_\phi + \theta_a \right) G^{\mu \nu}\tilde{G}_{\mu \nu}.
\end{align}
Such an interaction of the AD field with a gauge field can be generated through a fermion which is charged under the non-Abelian gauge group and which obtains a mass from the coupling with the AD field, see App.~\ref{app:Veff} for a concrete realization. We may neglect the fermion dynamics and simply use the interaction in Eq.~(\ref{eq:model1}) as long as the AD field value, and hence the fermion mass, is much larger than the temperature. This assumption will break down towards the end of the transfer; we will comment on the relevant fermion dynamics below.
The interaction~\eqref{eq:model1} breaks the axion shift symmetry and the $U(1)$ symmetry of the AD field down to the linear combination,
\begin{align}
    \frac{\dd}{\dd t}\left[ R^3 \left( \dot{\theta}_\phi r^2 -  \dot{\theta}_a f^2\right) \right] = 0 \,, 
\end{align}
allowing the charge of the AD field to be transferred to the axion field.
Here $R(t)$ denotes the scale factor of the Friedmann Robertson Walker metric.
The total comoving charge $R^3 q$ remains constant, with the physical charge
\begin{align}
    q = q_\phi - q_a = \dot{\theta}_\phi r^2 - \dot{\theta}_af^2 = m r^2 - \dot{\theta}_af^2
\end{align}
decreasing as $q \propto R^{-3}$ due to cosmic expansion, where in the last equality we have used $\dot{\theta}_\phi = m$.%
\footnote{This equality follows from the AD equation of motion which permits a stationary solution $\dot r = 0$ for $\dot \theta_\phi = m$ with equal kinetic and potential energy according to the virial theorem. 
}

\begin{figure}
\centering
 \includegraphics[width = 0.5 \textwidth]{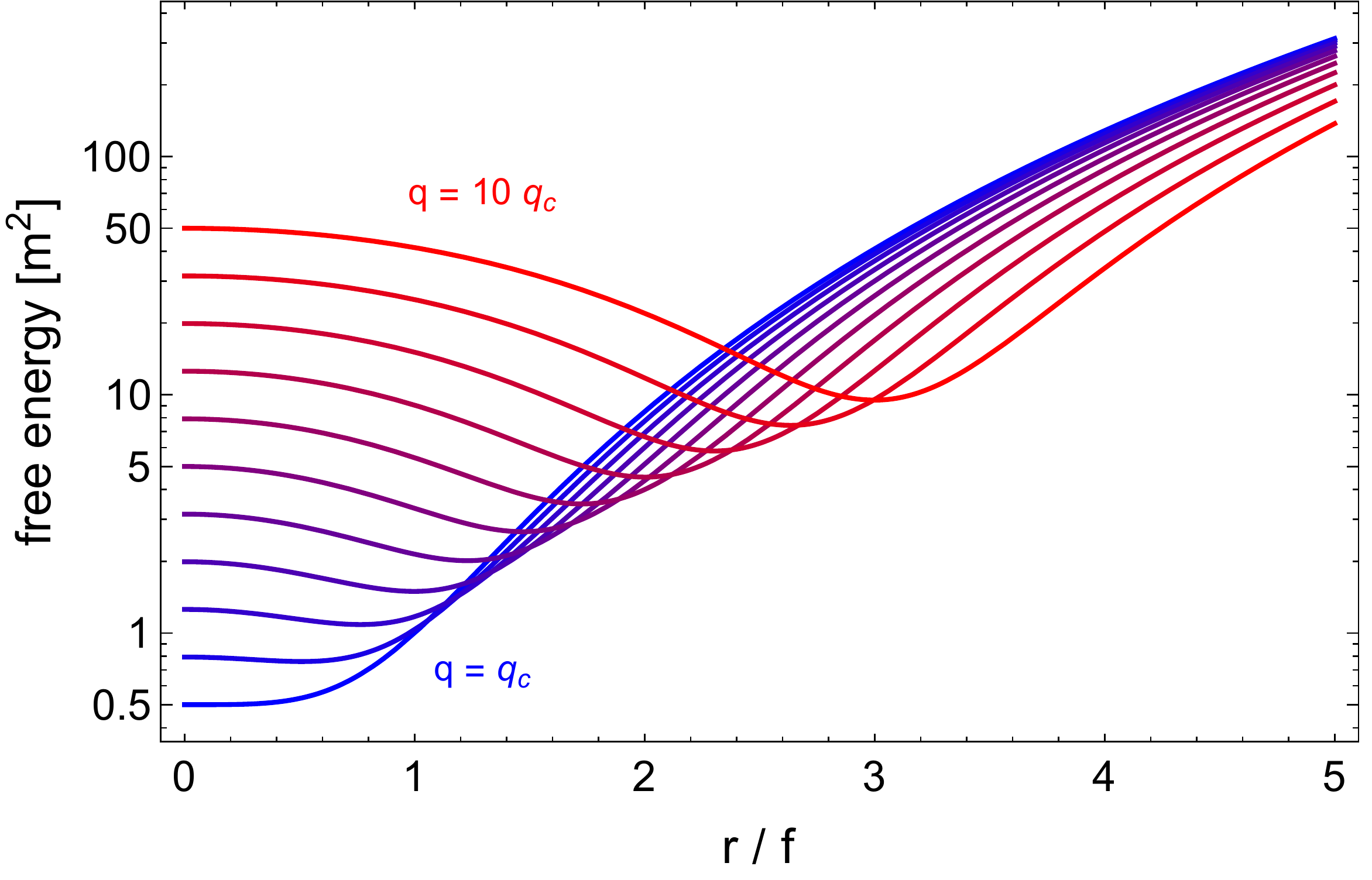}
 \caption{Free energy~\eqref{eq:energy} for different values of the total charge $q$, leading to a finite vev for the AD field for $q$ above the critical value $q_c$. For thermal corrections, see Sec.~\ref{sec:tunneling}.}
 \label{fig:FreeEnergyVac}
\end{figure}

\paragraph{Free energy.} Assuming that the charge transfer is efficient, whether the charge is dominantly stored in the AD or axion fields is determined so that the thermodynamic free energy density of the system is minimized. The contribution of a coherent rotation to the free energy density is simply the energy density.\footnote{
The change in (Helmholtz) free energy corresponds to the maximal amount of work a thermodynamic system can perform at a constant temperature. It can be obtained from the canonical partition function, which assumes a constant temperature and the number of particles. In practice, it is simpler to start from the grand canonical partition function, which assumes a constant temperature and chemical potential, and then obtain the Helmholtz free energy through a Legendre transformation. See App.~\ref{app:Veff} for details.
}
The AD field should have a vanishing ellipticity to minimize the energy density.
Using the charge conservation, the (free) energy density can then be expressed as a function of the axion velocity $\dot{\theta}_a$ or as a function of the AD field radius $r$,
\begin{align}
   \rho = m^2 r^2 + \frac{1}{2} \dot{\theta}_a^2 f^2 & =  m q + m \dot{\theta}_af^2 + \frac{1}{2}\dot{\theta}_a^2 f^2 \,, \nonumber \\
  & = m^2 f^2 \left[ \frac{q^2}{2 m^2 f^4} + \left(1 - \frac{q}{m f^2} \right) \frac{r^2}{f^2} + \frac{r^4}{2 f^4} \right] \,,
    \label{eq:energy}
\end{align}
which is minimized at $\dot{\theta}_a = - m $. The radius of the AD field at this equilibrium point is
\begin{align}
    r_{\rm eq}^2 = r_0^2 -f^2,
\end{align}
where $r_0= \sqrt{q/m}$ is the radius of the AD field when all of the charge $q$ is stored in the AD field. If $q \gg m f^2$ ($r_0 \gg f$), the charge remains dominantly stored in the AD field. 
Since cosmic expansion implies $r_0 \propto R^{-3/2}$, 
even if $r_0 \gg f$ initially, cosmic expansion will dilute the charge in the AD field, reducing $r_0$ to any critical value $r_c$, triggering the charge transfer to the axion field.
Once $q$ ($r_0$) is close to $m f^2$ ($f$), the larger fraction of the charge is transferred to the axion field, and for $q \rightarrow mf^2 \equiv q_c$ ($r_0 \rightarrow f \equiv r_c$), almost all charge is ends up in the axion field as $r_\text{eq} \rightarrow 0$, see Fig.~\ref{fig:FreeEnergyVac}.

For $q < q_c$, $r_{\rm eq}^2$ formally becomes negative, signaling the breakdown of our computation. In fact, as $q$ approaches $q_c$, the radius of the AD field decreases.
The mass of the fermion charged under $G$ and coupling to the AD field, responsible for the quantum anomaly in Eq.~(\ref{eq:model1}), is now much smaller than the temperature and hence this fermion can no longer be integrated out. The chiral asymmetry of the fermions should hence be also taken into account in the conservation law. Also, the charge asymmetry of the AD field should be no longer interpreted as a coherent rotation, but is rather given by particle-antiparticle asymmetry. The asymmetries of the fermion and the AD field are, however, of the order of $\dot{\theta}_aT^2$,
which is much smaller than the charge in the axion rotation. With this, we can estimate the velocity of the axion field from charge conservation,
\begin{align}
    \dot{\theta}_a = - \frac{q}{f^2} = - m \frac{r_0^2}{f^2}. 
\end{align}
See Sec.~\ref{sec:tunneling} for a discussion of the transition region $q \sim q_c$ accounting for thermal corrections.

In the above discussion, we implicitly assumed that the thermal bath is ``large", namely, that the back-reaction of the AD and axion fields to the thermal bath is negligible. In this picture, the equilibrium state minimizes the combined free energy of the AD and axion fields. One the other hand, when the energy of the AD field is comparable to the radiation energy, we must take into account the backreaction and treat the thermal bath together with the AD and axion fields as an isolated system. In this picture, we should rather maximise the entropy of the system, but the equilibrium state is still given by minimizing the energy of the AD and axion fields, since that is the state with the maximal temperature of the thermal bath and hence with the maximal entropy.

\paragraph{Transfer rate.} We have so far implicitly assumed that the charge transfer rate is larger than the Hubble expansion rate. Let us now estimate this transfer rate, see also App.~\ref{app:Boltzmann} for numerical results. The Boltzmann equations governing the transfer are
\begin{align}
    \frac{\dd}{\dd t} \left(\dot{\theta}_\phi r^2\right) = - \gamma_s T^2 \left(\dot{\theta}_\phi + \dot{\theta}_a\right) - 3 H \dot\theta_\phi r^2, \nonumber \\
    \frac{\dd}{\dd t} \left(\dot{\theta}_a f^2 \right) = - \gamma_s T^2 \left(\dot{\theta}_\phi + \dot{\theta}_a\right)  - 3 H \dot\theta_a f^2.
    \label{eq:Boltzmann1}
\end{align}
Here the first terms encode the charge transfer due to sphaleron processes, where $\gamma_s = c_\text{sph} \,  \alpha_G^4 T $ is the sphaleron transition rate with e.g.\ $c_\text{sph} \simeq 100$ for SU(3) sphalerons in the SM and $\alpha_G = g^2/(4 \pi)$ denoting the gauge coupling constant, and the second terms account for the cosmic expansion. 
At the equilibrium, $\tfrac{\dd}{\dd t}(R^3 \dot \theta_a f^2)= 0$, this yields $\dot{\theta}_a = - \dot{\theta}_\phi = -m$, confirming the computation based on the free energy.

For $r_0 > f$, starting from the initial condition $\dot{\theta}_a=0$, the velocity of the axion field approaches the equilibrium value $- \dot{\theta}_\phi$ sourced by the term $\gamma_s T^2 \dot{\theta}_\phi/f^2$. 
The rate of the transfer of comoving charge from the AD to axion fields is therefore given by
\begin{align}
    \Gamma (r_0 >f) = - \frac{\frac{\dd}{\dd t}( q_a R^3)}{(q_a - q_a^\text{eq})R^3} = \frac{\gamma_s T^2 (\dot \theta_\phi + \dot \theta_a)}{f^2(\dot \theta_a + m)} 
    = \frac{T^2}{f^2} \gamma_s \qquad \text{with} \quad  q_a^\text{eq} = - m f^2 ,
    \label{eq:Gamma0}
\end{align}
which is suppressed by a factor $T^2/f^2$ in comparison with the sphaleron transition rate $\gamma_s$.
Note that in this regime, only a very small fraction of the initial AD charge is transferred to the axion, $\dot q_\phi / q_\phi \ll \dot q_a / q_a$.

For $r_0 < f$, charge conservation implies the equilibrium value $\dot{\theta}_a = - m r_0^2 /f^2$. The rate is therefore
\begin{align}
    \Gamma (r_0 < f) =  \frac{\frac{\dd}{\dd t}(q_a R^3)}{q_a^\text{eq} R^3} \simeq \frac{- \gamma_s T^2 \dot \theta_\phi}{- m r_0^2} 
    = \frac{T^2}{r_0^2} \gamma_s \qquad \text{with} \quad  q_a^\text{eq} = - m r_0^2 ,
    \label{eq:Gamma1}
\end{align}
which is enhanced compared to Eq.~\eqref{eq:Gamma0} by a factor of $f^2/r_0^2$. 
In this regime, the charge density in the AD field is rapidly decreasing, $\dot q_\phi / q_\phi \sim \dot r/r  \sim \gamma_s T^2/r_0^2$. 
Note that growth rate of the axion charge in Eq.~\eqref{eq:Gamma1} is the same as this depletion rate for the AD field. This can be understood from the observation that once the AD field is depleted, the charge conservation requires that the depleted charge go to the axion field. In this picture, $r_0 < f$ is crucial, so that at the equilibrium the AD field loses almost all of its charge density. Note that we have assumed $r_0 >T$. Otherwise, the coherent picture of the AD field is not applicable and the transfer rate is simply given by $\gamma_s$. 

To determine if the charge transfer is efficient, we compare these comoving charge transfer rates with the Hubble expansion rate. For $r_0 > f$, the transfer rate decreases in proportion to $T^3$, faster than the Hubble expansion rate does. If the equilibrium was not reached initially, it is never reached when $r_0 >f$.
Even if the equilibrium is reached initially, the decoupling may occur later. 
For these cases, the axion velocity at a given temperature is given by 
\begin{align}
    \dot{\theta}_a \simeq -m \times \text{min}\left[1, \frac{\Gamma(r_0 > f)}{3  H}\right] = -  \frac{\gamma_s T^2 m}{3 f^2 H} \propto T \quad \text{for $\Gamma \ll H$} .
    \label{eq:theta_a_inefficient_transfer}
\end{align}
Note that the axion velocity decreases only linearly with $T$, slower than what would be caused by the Hubble friction ($\propto T^3$), so the axion velocity at a given temperature is dominated by the charge transfer at that temperature and rather insensitive to previous charge transfer at higher temperatures.
Once $r_0$ drops below $f$, the transfer rate remains constant (recall $r_0^2 \propto q \propto T^3$) and thus eventually dominates over the Hubble expansion rate. Even if the axion field has not yet reached the equilibrium value when $r_0 \sim f$ (for which the AD field would collapse toward the origin given an efficient charge transfer),
the equilibrium is eventually reached, and almost all charge is transferred into the axion field. 
In summary, cosmic expansion acts towards removing the metastable minimum of vacuum potential of the AD field at finite field value $r_\text{eq}$ and eventually ensures an efficient transfer of the AD charge to the axion field.

\paragraph{Grand canonical ensemble.} Before closing this subsection, we compute the equilibrium value of $r$ using the grand canonical ensemble with the chemical potential $\mu$ associated with the total charge $q$ and obtaining the free energy density by a Legendre transformation. This framework is reviewed in Appendix~\ref{app:Veff} and used in the remainder of this paper.
The effective potential from the rotation of the AD field and axion fields under a fixed chemical potential $\mu$ is\footnote{
  Here the subscript $\mu$ is a reminder that this effective potential is obtained under a fixed $\mu$.
}
\begin{align}
    V_{\mu,\text{eff}} (r, \mu) = \frac{1}{2} m^2 r^2 - \frac{1}{2}\mu^2 (r^2 + f^2),
\end{align}
which is related to the thermodynamic pressure $p$ after extremizing it with respect to $r$ (see Appendix~\ref{app:Veff}).
The relation between $q$ and $\mu$ is
\begin{align}
\label{eq:qmu_toy1}
    q = - \frac{\partial V_{\mu, \text{eff}} (r,\mu)}{\partial \mu} = \mu (r^2 + f^2).
\end{align}
In our context, we would like to minimize the free energy for a fixed $q$ rather than a fixed $\mu$ since we do not have the bath for the charge $q$.
Hence, the effective potential in our context is obtained through the Legendre transformation,
\begin{align}
    V_{\rm eff}(r,q) 
    = V_{\mu, \text{eff}} (r, \mu) + \mu q
    =  \frac{1}{2} m^2 r^2 + \frac{q^2}{2(r^2 + f^2)}.
    \label{eq:Veff1}
\end{align}
This is minimized at $r^2 = q/m-f^2$, implying $\mu = m$.
Using the relation (\ref{eq:qmu_toy1}) and $q = \dot{\theta}_\phi r^2 - \dot{\theta}_a f^2$, we obtain the equilibrium value $\dot{\theta}_\phi = -\dot{\theta}_a = m$.

\subsection{Charge transfer including charged fermions}
\label{sec:fermion}

\paragraph{Conserved charges and free energy.} We next consider an example that will illustrate an important condition on the symmetries of the system for the transfer to occur in the presence of additional (approximate) symmetries.
To the system discussed in the previous subsection, let us add a gauge charged Dirac fermion pair $\psi$ and $ \overline{\psi}$. If the Dirac fermion is massless, there are three conserved charges (up to cosmic expansion),
\begin{align}
\label{eq:charge2}
    \dot{\theta}_\phi r^2 - \dot{\theta}_a f^2 \equiv q (\neq 0)\,, \quad  \frac{1}{2}\left( q_\psi + q_{\bar{\psi}} \right)- \dot{\theta}_a f^2 \equiv q_A (=0)\,, \quad q_\psi - q_{\bar{\psi}} \equiv q_B (=0),
\end{align}
where $q_\psi$ and $q_{\bar {\psi}}$ are particle-antiparticle asymmetry of $\psi$ and $\bar{\psi}$, respectively, and we have chosen the initial condition so that the latter two conserved charges vanish. The associated chemical potentials are $\mu$, $\mu_A$, and $\mu_B$, respectively.
The chemical potential-dependent part of the effective potential is (see App.~\ref{app:Veff})
\begin{align}
    - V_{\mu, \text{eff}} (r, T, \bm{\mu})
    \supset \frac{1}{2} \mu^2 r^2 + \frac{1}{2} \qty(\mu+ \mu_A)^2 f^2 + \frac{d_\psi}{6} \qty(\frac{1}{4}\mu_A^2 + \mu_B^2) T^2,
\end{align}
where $\bm{\mu}$ collectively denotes the chemical potentials $\bm{\mu} = (\mu, \mu_A, \mu_B)$, and $d_\psi$ is the dimension of the gauge representation of $\psi$, i.e., $d_\psi = 1$ for a gauge-singlet Dirac fermion.
The relation between the conserved charges and chemical potentials is
\begin{align}
    q = 
    - \frac{V_{\mu, \text{eff}} (r, T, \bm{\mu})}{\partial \mu}
    & = \qty(\mu + \mu_A)f^2 + \mu r^2, \nonumber \\
    q_A =
    - \frac{V_{\mu, \text{eff}} (r, T, \bm{\mu})}{\partial \mu_A}
    & = \qty(\mu + \mu_A)f^2 + \frac{d_\psi}{12} \mu_A T^2, \nonumber \\
    q_B =
    - \frac{V_{\mu, \text{eff}} (r, T, \bm{\mu})}{\partial \mu_B}
    & = \frac{d_\psi}{3} \mu_B T^2. 
\end{align}
Using $q_A = q_B = 0$, we obtain
\begin{align}
    \mu &= \frac{q}{r^2} \frac{1 + d _\psi T^2/(12 f^2)}{1 + d_\psi T^2/(12 r^2) + d_\psi T^2/(12 f^2)}, \nonumber \\
    \mu_A  &= -\frac{q}{r^2} \frac{1}{1 + d_\psi T^2/(12 r^2) + d_\psi T^2/(12 f^2)}, \nonumber \\
    \mu_B &= 0 .
\end{align}
The fixed-$q$ effective potential from the AD and axion rotations and the asymmetry of $\psi$ is obtained as
\begin{align}
    V_{\rm eff}(r,T,q) 
    &= V_{\mu, \text{eff}} (r, T, \bm{\mu})_{q_{A,B} = 0} + \mu q \nonumber \\
    &=  \frac{1}{2}m^2 r^2 + \frac{q^2}{2r^2} \frac{1 + d _\psi T^2/(12 f^2)}{1 + d_\psi T^2/(12 r^2) + d_\psi T^2/(12 f^2)} \simeq \frac{1}{2}m^2 r^2 + \frac{q^2}{2 r^2},
\end{align}
where we assume $r,f \gg T$ in the last inequality.
The effective potential is minimized at $r^2 \simeq q/m = r_0^2$.
At the equilibrium, the charge of the axion rotation is
\begin{align}
    \qty(\mu + \mu_A)f^2 = \frac{d_\psi}{12} m T^2 \frac{1}{1 + d_\psi T^2/(12 f^2)} \simeq \frac{d_\psi}{12} m T^2  ,
\end{align}
where we used $f \gg T$ in the  second equality.

In contrast to the situation discussed in Sec.~\ref{sec:gauge}, the axion obtains a subdominant fraction of the charge as long as $q > m T^2 $ (i.e., $r_0 >T$)  even if $r_0 \lesssim f$. This is due to the second conservation law in Eq.~(\ref{eq:charge2}); for the axion rotation to obtain a large charge, the charge asymmetry of the fermion must be also large, leading to a large free energy. If the Dirac fermion has a non-zero Dirac mass, the conservation law is violated and the axion field can obtain a large charge. In general, in order for the axion field to obtain most of the charge, all symmetries that would require large charge asymmetry of fermions must be violated. For example, any chiral symmetry with $G$ anomaly must be explicitly broken.

\paragraph{Transfer rate.} Let us estimate the transfer rate including the mass term $m_\psi \psi \overline{\psi}$. The Boltzmann equation is given by
\begin{align}
    \frac{\dd}{\dd t} \left(\dot{\theta}_\phi r^2\right) =& - \gamma_s  \left(\dot{\theta}_\phi T^2 + \dot{\theta}_a T^2 + q_\chi \right) - 3 H \dot \theta_\phi r^2, \nonumber \\
    \frac{\dd}{\dd t} \left(\dot{\theta}_a f^2 \right) =& - \gamma_s  \left(\dot{\theta}_\phi T^2 + \dot{\theta}_a T^2 + q_\chi \right) - 3 H \dot \theta_a f^2, \nonumber \\
    \frac{\dd}{\dd t}q_\chi =&  - 2 \gamma_s  \left(\dot{\theta}_\phi T^2 + \dot{\theta}_a T^2 + q_\chi \right) - \gamma_\chi q_\chi -3 H q_\chi,  \label{eq:boltzmann_mass}
\end{align}
where $\gamma_\chi$ is the chiral symmetry breaking rate $\sim \alpha_G m_\psi^2/T$ and $q_\chi = q_\psi + q_{\overline{\psi}}$.
Starting from the initial condition $\dot{\theta}_a T^2=q_\chi =0$ and $\dot{\theta}_\phi = m$, the fermion charge $q_\chi$ first reaches the quasi-equilibrium value determined by $\dd (q_\chi R^3)/\dd t \simeq \dd q_\chi/\dd t = 0$, 
\begin{align}
  q_{A,{\rm eq}} = - \frac{2 \gamma_s}{\gamma_\chi + 2 \gamma_s} \left(\dot{\theta}_\phi + \dot{\theta}_a \right)T^2,
    \label{eq:boltzmann_mass-eq}
\end{align}
with a rate 
\begin{align}
\label{eq:rate_qA}
   \frac{\frac{\dd}{\dd t} (R^3 q_\chi)}{R^3 q_\chi^\text{eq}} \simeq  - \frac{2 \gamma_s \qty(\dot{\theta}_\phi+\dot{\theta}_a ) T^2}{q_{A,{\rm eq}}} = \gamma_\chi + 2 \gamma_s,
\end{align}
for $q_\chi \ll q_\chi^\text{eq}$.
As we will see shortly, the system reaches the true equilibrium state with a rate suppressed by $T^2/ {\rm min}(r_0^2,f^2)$ and the assumption of $q_\chi$ reaching the quasi-equilibrium value is consistent. Putting $q_{A,{\rm eq}}$ to Eq.~(\ref{eq:boltzmann_mass-eq}), we obtain the Boltzmann equation of $\dot{\theta}_\phi r^2$ and $\dot{\theta}_a f^2$ at this quasi-equilibrium,
\begin{align}
    \frac{\dd}{\dd t} \left({R^3} \dot{\theta}_\phi r^2\right) = \frac{\dd}{\dd t} \left({R^3} \dot{\theta}_a f^2 \right) = - \frac{\gamma_s \gamma_\chi}{\gamma_\chi + 2 \gamma_s}\qty(\dot{\theta}_\phi+\dot{\theta}_a ) T^2 {R^3}.
\end{align}
For $\gamma_\chi \gg \gamma_s$, we recover Eq.~\eqref{eq:Boltzmann1}, i.e., an efficient charge between the AD field and the axion.

Applying the same discussion as Sec.~\ref{sec:gauge}, we obtain the transfer rate from the AD to axion field
\begin{align}
    \Gamma = \frac{\gamma_s \gamma_\chi}{\gamma_\chi + 2 \gamma_s}  \frac{T^2}{f^2 + r_0^2} \simeq {\rm min}\qty(\gamma_s,\gamma_\chi) \times  \frac{T^2}{{\rm min}\qty(f^2, r_0^2)}.
\end{align}
As anticipated, the rate is much smaller than the rate in Eq.~(\ref{eq:rate_qA}) and the assumption of $q_\chi$ first reaching the quasi-equilibrium value is justified.
For $\gamma_\chi < \gamma_s$, the transfer rate $\Gamma$ decreases slower than $H$ does even if $r_0 > f$. Thus, even if the transfer is not effective at high temperatures, it can be effective at low temperatures.

\section{Thermal potential and charge transfer by thermal fluctuations}
\label{sec:tunneling}

In the previous section, we have shown how charge can be transferred from an AD field to an axion field assuming a zero-temperature quadratic potential for the AD field. In realistic setups, the AD field couples to the thermal bath and may obtain a non-negligible thermal potential. In this section, we discuss the effect of thermal corrections to the potential and show that the charge transfer may involve thermal tunneling.

\subsection{Effective potential for the AD field}
\label{sec:f w/o axion}

\paragraph{Setup.} Let us first discuss the dynamics of the AD field taking into account thermal corrections, but for now ignoring the charge transfer to the axion field. We consider the following exemplary setup:
\begin{itemize}
    \item The AD field spontaneously breaks a  gauge symmetry, giving a mass $g r$ to the gauge bosons.
    \item The AD field moreover couples to fermions $\psi$ in a thermal bath (whose mass does not depend on the AD field), such that the charge of the AD field can be transferred into a particle-antiparticle asymmetry of these particles. 
\end{itemize}
The conserved charge is
\begin{align}
    q = q_\phi + q_\psi
\end{align}
with the associated chemical potential $\mu$. Without loss of generality, we take $q >0$. 
In realistic setups, there typically also exist fermions which contribute to the conserved charge $q$ and obtain masses from the AD field. We analyse such a case in App.~\ref{app:Veff} and find that the qualitative behavior of the potential does not change.

\paragraph{Local minima of the effective potential.} Following the formulation described in Appendix~\ref{app:Veff},  parts of the effective potential depending on the AD field radius $r$ and the total charge $q$ are given by 
\begin{align}
    V_{\rm eff}(r,T,q)= &V_0(r)+ V_{g}(r,T) + V_q(r,T,q), \nonumber \\
    V_0(r) =& \frac{1}{2}m ^2 r^2,~~ \quad 
    V_{g}(r,T)= 2 N_g \frac{T^4}{2\pi^2} \int \dd x x^2 {\rm log}\left(1- e^{-\sqrt{x^2 + g^2 r^2/T^2}}\right), \nonumber \\
    V_q(r,T,q) = & \frac{1}{12} \mu(r,T,q)^2 \qty(T^2 + 6 r^2) + \frac{1}{8\pi^2} \mu(r,T,q)^4,\\
    \label{eq:q_mu}
    q =&  \frac{1}{6} \mu(r,T,q) \qty(T^2 + 6 r^2) + \frac{1}{6\pi^2} \mu(r,T,q)^3.
\end{align}
Here $V_0$ is the vacuum potential of the AD field and $V_{g}$ is the free energy of $N_g$ gauge fields which obtain a mass from the AD field.  (To be precise, a resummation to include the thermal mass of the gauge boson is necessary, but we find that this does not change the qualitative behavior.) $V_q$ is the $q$-dependent part of the free-energy contribution associated with the asymmetry of $\psi$ and the rotation of the AD field. The chemical potential $\mu$ should be determined as a function of $(r,T,q)$ according to Eq.~(\ref{eq:q_mu}).
The fluctuations of the AD field also contribute to the free energy, but  their inclusion does not change the qualitative behaviors of the free energy; see App.~\ref{app:Veff}.

\begin{figure}[t]
\centering
 \includegraphics[width = 0.47 \textwidth]{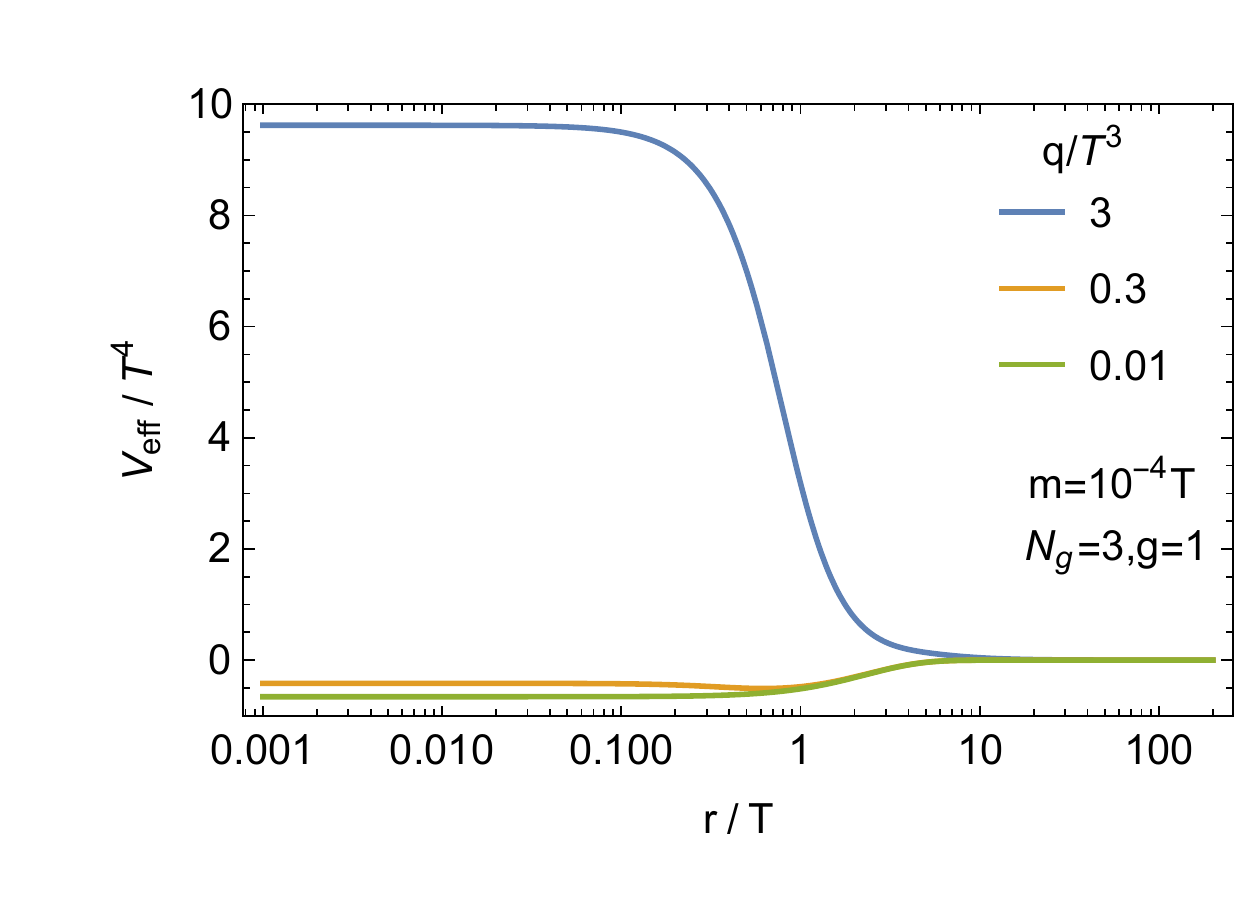} \hfill
  \includegraphics[width = 0.47 \textwidth]{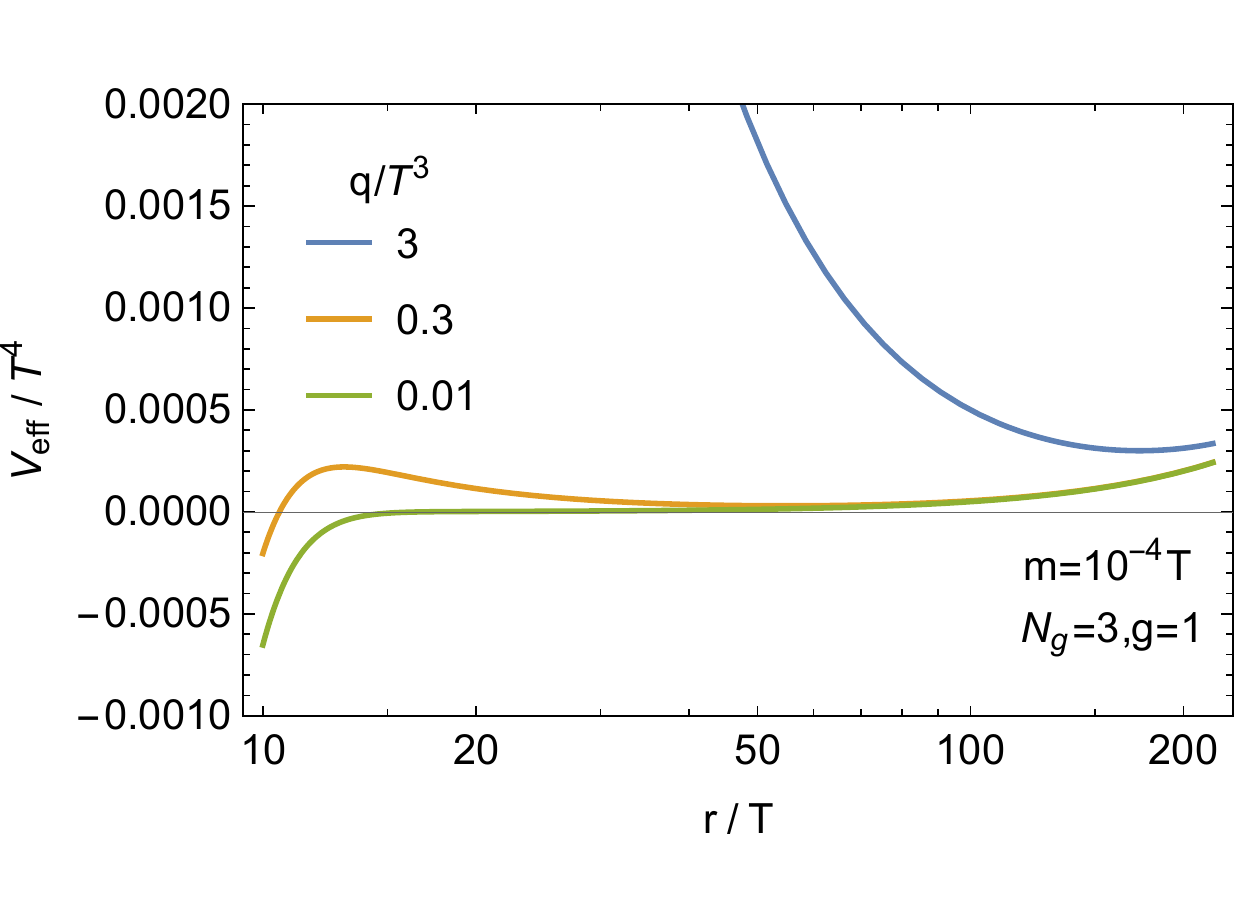} 
 \caption{Effective potential including thermal contributions for the case without a charge transfer to the axion as a function of the radius of the AD field $r$ for several total charges $q$. The left panel shows a global picture, while the right panel focuses on the potential around the minimum at $r_0$.}
 \label{fig:free_energy_wo_axion}
\end{figure}

In Fig.~\ref{fig:free_energy_wo_axion}, we show the effective potential as a function of $r$ for several values of $q$. Here we take $m=10^{-4}T$. At large $q > T^3$, the free-energy density has a unique minimum at $r \simeq  \sqrt{q/m}= r_0$. As $q$ approaches $T^3$, $r_0$ becomes a local minimum and a global minimum appears at a smaller $r$. For $q \ll T^3$, $r_0$ is no longer a local minimum and the unique minimum is at $r=0$. This behaviour is due to several competing effects as we explain below. Large values of the AD charge imply a centrifugal force which creates an a local minimum for the radial component of the AD field at finite $r$.
This is completely analogous to our observation in Sec.~\ref{sec:gauge} before the charge transfer to the axion becomes efficient.
On the other hand, thermal masses for the gauge bosons drive a restoration of the gauge symmetry at high $T$, thus pushing the order parameter $r$ of the spontaneous gauge symmetry breaking to zero. Consequently we expect a global minimum at $r = 0$ to develop once $q/T^3 \lesssim 1$~\cite{Laine:1998rg}.

Let us understand this behavior analytically.
The first and second derivatives of $V_{\rm eff}$ are
\begin{align}
\label{eq:derivative}
\frac{\partial V_{\rm eff}}{\partial r} = & \frac{\partial \qty(V_0 + V_{g})}{\partial r} - \mu^2 r,\nonumber\\
\frac{\partial^2 V_{\rm eff}}{\partial r^2} = & \frac{\partial^2 \qty(V_0 + V_{g})}{\partial r^2}- \mu^2- 2 \mu r \frac{\partial \mu}{\partial r}.
\end{align}
Thus, $V_{\rm eff}$ has extrema at $r=0$ and at non-zero $r$ which satisfy $m^2 + (\partial V_g/\partial r) / r =\mu(r,T,q)^2$. When $\partial V_g/\partial r$ is negligible, in the limit $m \ll T$, the latter has a solution $r^2 \simeq q/m-T^2/6 \equiv r_{\rm eq}^2$. So when $q\gg m T^2$, there is an extrema at $r= \sqrt{q/m} = r_0$.

The second derivative at $r=0$ is
\begin{align}
\label{eq:2nd_derivative0}
\left(\frac{\partial^2 V_{\rm eff}}{\partial r^2}\right)_{r=0} = m^2 + \frac{g^2 N_g}{6} T^2 - \mu^2.
\end{align}
By using the relation between $q$ and $\mu$ at $r=0$, one can show that if $q\gtrsim 0.07 g\sqrt{N_g} T^3$, the third term dominates and the second derivative is negative; $r=0$ is a maximum. If $q\lesssim 0.07 g\sqrt{N_g} T^3$, $r=0$ is an minimum. This can be intuitively understood in the following way. Having $r=0$ requires that all charge be stored in the chiral asymmetry of the fermions and hence the chemical potential is fixed. For large $q$, the chemical potential is so large that the thermal mass from the gauge bosons cannot keep the AD field at the origin and hence a BEC forms.

To obtain the second derivative at $r= r_{\rm eq}$, we need $\partial \mu/\partial r$. This can be obtained by differentiating Eq.~(\ref{eq:q_mu}) with respect to $r$. This yields
\begin{align}
\left(\frac{\partial^2 V_{\rm eff}}{\partial r^2}\right)_{r=r_{\rm eq}} = \left[ V_0'' + V_g'' - \frac{V_0'+V_g'}{r} + \frac{4 r \qty(V_0'+V_g')}{r^2 + \frac{1}{6}T^2 + \frac{1}{24\pi^2 r}\qty(V_0'+V_g') }\right]_{r= r_{\rm eq}},
\end{align}
where the primes denote the derivative with respect to $r$. For $q \gg mT^2$, $r_{\rm eq}^2\simeq q/m \gg T^2$. For such large $r$, $V_g'$ is exponentially suppressed and
\begin{align}
\left(\frac{\partial^2 V_{\rm eff}}{\partial r^2}\right)_{r=r_{\rm eq}} \simeq 4 m^2 >0, 
\end{align}
so $r_{\rm eq}$ is a minimum. As $q$ approaches $m T^2$, $r_{\rm eq}$ approaches $T$ and $V_g$ is not negligible, and the minimum at $r_{\rm eq}$ eventually disappears.
The non-zero value of $r$ for large enough $q$ can be understood as a result of the centrifugal force provided by the rotation.

For $m T^2 \ll q \lesssim 0.07 g\sqrt{N_g} T^3 $, there are two minima at $r=0$ and $r_{\rm eq}$. The free energies at these two minima are
\begin{align}
V_{\rm eff} \qty(0,T,q) \simeq&
3 q^{4/3} \,, \nonumber\\ 
    V_{\rm eff}\qty(r_{0},T,q) \simeq & m q + \frac{\pi^2 N_g}{15} T^4 \simeq \frac{\pi^2 N_g}{15} T^4 \,.
\label{eq:Veff_comparison}
\end{align}
Therefore, for $q \lesssim 0.3 N_g^{3/4} T^3$, the minimum at $r=0$ is the global minimum and that at $r_{\rm eq}$ is a local minimum. The inequality is indeed satisfied for $q \lesssim 0.07 g\sqrt{N_g} T^3$ with perturbative $g$.

To sum up, if $q\gtrsim 0.07 g\sqrt{N_g} T^3$, there is a unique minimum at $r^2 = q/m$ and if $m T^2 \ll q\lesssim 0.07 g\sqrt{N_g} T^3$, there is a global minimum at $r=0$ and a local minimum at $r^2 = q/m$. The local minimum disappears as $q$ approaches $m T^2$.

\paragraph{Additional contributions to the thermal potential.} Before closing this subsection, we point out another possible contribution to thermal potential of $r$ that appears in realistic setups and can destabilize the local minimum at $r^2 = q/m$. After integrating out the heavy degrees of freedom that obtain masses from the AD field, some of the couplings of the low energy effective theory may depend on the AD field value. Consequently, the corrections to the free energy arising from these couplings depend on the AD field value, generating additional thermal potential of the AD field. Let us consider an exemplary situation where the gauge coupling constant depends on the AD field and a thermal potential $\sim \alpha^2 T^4 {\rm ln}(r^2/T^2)$ is generated~\cite{Anisimov:2000wx}. The potential gradient from this term exceeds the gradient arising from $V_0+ V_q$ at $r^2 = q/m$ if $q < \alpha^2 T^4/m$ (assuming $ T < q/T^3 (m/\alpha^2)$),
leading to a destabilization of the local minimum at $r^2 = q/m$.

\subsection{Effective potential including the axion}
\label{sec:f w axion}
We now include the axion field. As we will see, in this case the origin $r=0$ becomes the absolute minimum more easily than in the case discussed above, which  neglected the charge transfer to the axion field. This is in line with the discussion in Sec.~\ref{sec:gauge}, where we saw that the charge transfer to the axion drives the local minimum for the AD field to zero.
The total conserved charge is now
\begin{align}
    q = q_\phi + q_\psi - q_{a}
\end{align}
with the associated chemical potential $\mu$.

\paragraph{Local minima of the effective potential.} The effective potential is given by (see App.~\ref{app:Veff})
\begin{align}
    V_{\rm eff} \qty(r,T,q)= &V_0(r)+ V_{g}(r,T) + V_{q,a}(r,T,q), \nonumber \\ 
    V_{q,a} \qty(r,T,q) = & \frac{1}{12} \mu(r,T,q)^2 \qty(T^2 + 6 r^2 + 6 f^2) + \frac{1}{8\pi^2} \mu(r,T,q)^4,\\
    \label{eq:qmu_axion}
    q =&  \frac{1}{6} \mu(r,T,q) \qty(T^2 + 6 r^2 + 6 f^2) + \frac{1}{6\pi^2} \mu(r,T,q)^3.
\end{align}
The expression is greatly simplified when $q \ll T f^2 $, for which the $\mu^4$ term in $V_{q,a}$ and the $\mu^3$ term in $q$ are always negligible. Then $V_{q,a}$ is given by
\begin{align}
    V_{q,a} \qty(r,T,q) \simeq \frac{3q^2}{6 \qty(r^2+ f^2) + T^2}.
\end{align}

\begin{figure}[t]
\centering
 \includegraphics[width = 0.48 \textwidth]{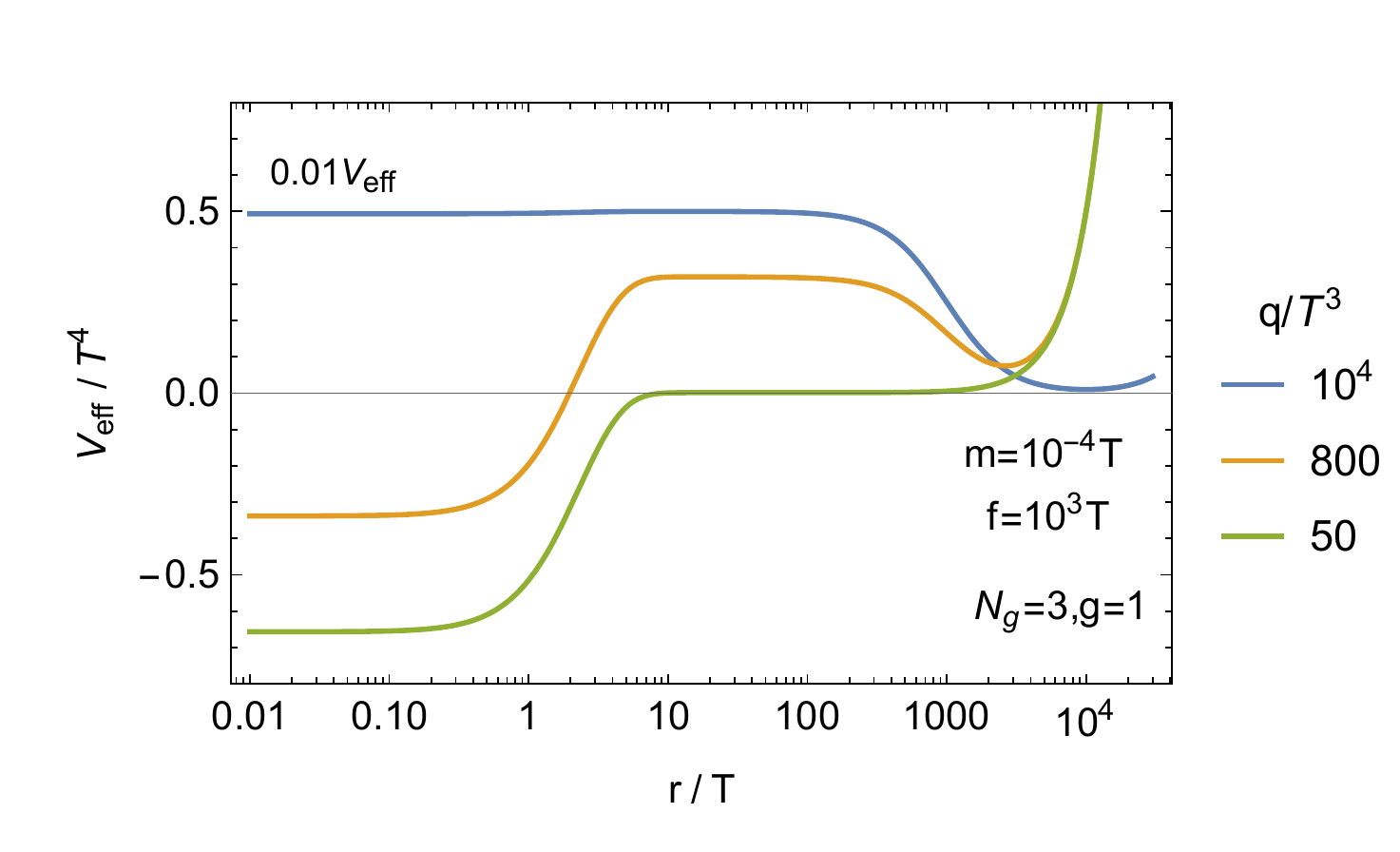} \quad
 \includegraphics[width = 0.48 \textwidth]{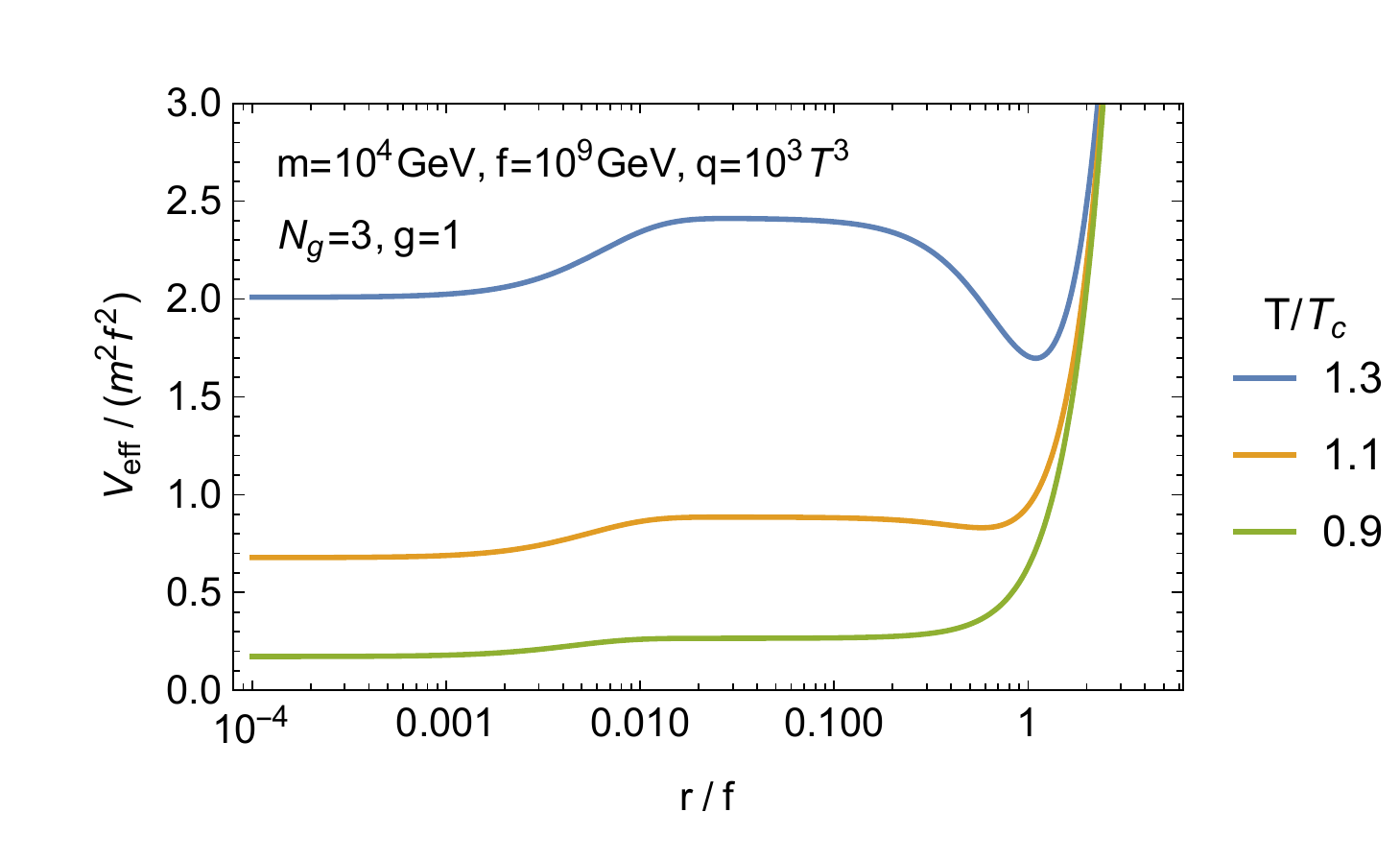} 
 \caption{Effective potential including thermal contributions for the case with  charge transfer to the axion as a function of the radius of the AD field $r$ for several total charges $q$.}
 \label{fig:free_energy_w_axion}
\end{figure}

In the upper panel of Fig.~\ref{fig:free_energy_w_axion}, we show the effective potential as a function of $r$ for several representative values of $q$. When $q$ is very large, there is a unique minimum at $r=r_0$. As $q$ decreases, even if $q > T^3$, $r_0$ becomes only a local minimum and the global minimum is at $r=0$. As $q$ approaches $q_c$, the local minimum approaches $0$ and eventually disappears. The lower panel shows the evolution of the effective potential for a fixed $m$, $f$, and $q/T^3$ around the time when the transfer completes. Here $T_c$ is the temperature at which $q=m f^2$. The minimum at $r_0$ becomes an local one as $T$ approaches $T_c$, and disappears for $T< T_c$.

Let us understand this behavior analytically. The first and the second derivatives of $V_{\rm eff}$ are given by Eq.~(\ref{eq:derivative}), and the second derivative at $r=0$ is given by Eq.~(\ref{eq:2nd_derivative0}). When $q \gg f^2 T$, Eq.~(\ref{eq:qmu_axion}) gives $\mu \qty(0,T,q) \simeq q/f^2  \gg T$. Then the second derivative of $V_{\rm eff}$ at $r=0$ is negative, so the origin cannot be a minimum. Instead, there is a unique minimum at $r$ such that 
$\mu \qty(r,q,T) \simeq m$, namely, $r^2\simeq q/m (\gg f^2)$.
When $q < f^2 T$, $\mu \qty(0,T,q)\simeq q/f^2  < T^2$, so the origin is a minimum. For $q > m f^2$, $r_{\rm eq}$ is also a minimum. The effective potentials at $r=0$ and $r_{\rm eq}$ are
\begin{align}
    V_{\rm eff} \qty(0,T,q) \simeq & \frac{q^2}{2 f^2}, \nonumber \\
    V_{\rm eff} \qty(r_{\rm eq},T,q) \simeq &  m q+ \frac{\pi^2 N_g}{15} T^4.
\end{align}
When $q > (2\pi^2 N_g/15)^{1/2} f T^2$, $r_{\rm eq}$ has a smaller effective potential and is the global minimum, while for $q < (2\pi^2 N_g/15)^{1/2} f T^2$, $r=0$ is the global minimum.
As $q$ approaches $mf^2$, $r_{\rm eq} = \sqrt{q/m-f^2}$ approaches $T$, and $V_g$ eventually destabilizes the local minimum at $r_{\rm eq}$.

If $q/T^3 > (2\pi^2 N_g/15)^{3/4}(f/m)^{1/2}$, $q$ drops below $mf^2$ before it drops below $(2\pi^2 N_g/15)^{1/2} f T^2$.
In this case, $r_{\rm eq}$ continues to be the global minimum until $q$ becomes very close to $m f^2$. 
For such a large $q/T^3$, the AD field dominates the energy density of the universe before $q$ reaches the critical value and kination domination occurs after the charge is transferred into the axion field, see Sec.~\ref{sec:pheno}.

\subsection{Phase transition including the thermal corrections}
\label{subsec:PT}

With the effective potentials we have shown above, we now discuss the evolution of the AD field and the axion field towards the end of the charge transfer around $q = q_c$. We assume $q/T^3 \gg1$ and $T \gg m$.

When $q >  (2\pi^2 N_g/15)^{1/2} f T^2$, the state with almost all charge in the rotation of the AD field has the least free-energy, and the rotation with $r^2 = q/m -f^2 \simeq q/m $ is absolutely stable.
As the temperature decreases, $q$ drops below $ (2\pi^2 N_g/15)^{1/2} f T^2$ and  $r^2  \simeq q/m$ becomes a local minimum, while the minimal free energy is  achieved by the state where almost all of charges are in the axion rotation and the AD field is trapped around the origin. A thermal transition from the local minimum to the absolute minimum can occur, but since $r^2 \simeq q/m \gg T^2$, we expect that the transition is suppressed and does not occur within the cosmological time scale.
As $q$ approaches $q_c = m f^2 $, $r^2= q/m-f^2$ approaches $T^2$, and the transition to the absolute minimum may become efficient. Note that at this point $q-q_c$ is much smaller than $q_c$, so almost all charges have been already transferred into the axion field before the thermal transition becomes effective.
The thermal transition may occur as a first order phase transition, which proceeds  via nucleation of bubbles~\cite{Coleman:1977py,Callan:1977pt,Linde:1981zj}, or as phase mixing, where the two minima are populated via creation of subcritical bubbles~\cite{Gleiser:1991rf,Dine:1992wr,Shiromizu:1995jv,Hiramatsu:2014uta}.

\paragraph{Uncertainties in the phase transition dynamics.} A full investigation of the transition dynamics is beyond the scope of the present paper. Instead we highlight only some differences from the usual phase transition problem:
First, the fields relevant for the phase transition rotate rather than just moving in the radial direction. The effect of the angular motion in field space and associated conserved charges must hence be taken into account. 
Second, the charge can be transferred from the AD field both to the axion field and the thermal bath. Whereas the transfer rate for the former is suppressed by $T^2/f^2$, the latter is only suppressed only by $T^2/r^2$, which is much larger than $T^2/f^2$ when $q$ is close to $q_c$. Also, the transfer involving the axion may require additional processes, further suppressing the transfer rate. For example, in the toy model discussed in Sec.~\ref{sec:fermion}, the transfer rate into the axion involves the chiral symmetry breaking by the fermion mass. 
This implies that the charge transfer during the tunneling process may involve only the AD field and the thermal bath, or could also involve the axion field. In the former case, we expect bubbles of true vacuum to form with $r = 0$ and $\dot \theta_a = m$ in a false vacuum with $r = r_\text{eq}$ and $\dot \theta_a = m$. In the latter case, bubbles are formed with $r = 0$ and $\dot \theta_a > m$, with the charge of the AD field transmitted to the rotation of the PQ field inside the bubble. Once these bubbles expand, more charge is converted from the AD field into the axion motion and/or the thermal bath. Again, a comparison between the two competing transfer rates is required to understand the dynamics of the bubble expansion. 
Finally, unlike the AD field, the axion field does not experience any phase transition during the tunneling; only the axion velocity changes, which is not an order parameter. This means that
sub-critical bubbles can be more important than the usual case; inside the sub-critical bubble, the AD field is at the origin while the axion field has a larger charge density than outside. As the sub-critical bubble collapses, the AD field will revert to large field values, but the excess of the axion charge inside the bubble can spread out as density waves.

\paragraph{Inhomogeneities in the axion field.} The first order phase transition or the phase mixing at the end of the transfer could in fact be phenomenologically interesting. After the first order phase transition or the phase mixing, the axion field will have large fluctuations, which can contribute to axion dark matter if they are not thermalized.%
\footnote{One may worry that the fluctuations may produce stable domain walls. We expect this is not the case, since the fluctuations are produced in sub-horizon modes.}
Whether or not thermalization occurs will depend on the length scale of the fluctuations, since the interaction of the axion with the thermal bath is suppressed for long-wavelength modes.
We also note that Q-balls may form when $q$ is close to $q_c$. For $q \gtrsim q_c$, $r$ is not much above $T$, for which the thermal potential becomes comparable to the vacuum potential. Since the thermal potential is flatter than the quadratic one, fluctuations around the rotating background grow and Q-balls are formed. This occurs only after the majority of the charges have been transferred into the axion field, so the basic picture of the transfer is not altered. However, the fluctuations associated with the Q-ball formation may have some phenomenological implications, such as the production of axion dark matter.
These observations motivate further studies on the dynamics of the transition, which are beyond the scope of the present paper. We will instead present a lower bound on the axion dark matter production based on the homogeneous axion component in Sec.~\ref{sec:DM}.

\section{Phenomenological applications}
\label{sec:pheno}

In this section, we discuss phenomenological applications of the charge transfer from the AD field to the axion field.

\subsection{Axion dark matter by kinetic misalignment \label{sec:DM}}

If the axion field receives a large enough charge, the axion field continues to rotate even after the Hubble expansion rate drops below the axion mass $m_a$ and the axion would begin to oscillate around the minimum. In this case, 
axions are produced by the kinetic misalignment mechanism~\cite{Co:2019jts}, where the kinetic energy of the axion rotation is transferred into axion dark matter density. The resulting number density of the axion $n_a$ is as large as the charge density $q_a$, so
\begin{align}
    \frac{\rho_a}{s} \simeq m_a Y_a = 0.4~{\rm eV} \frac{m_a}{\rm meV} \frac{Y_a}{400} = 0.4~{\rm eV} \frac{10^9~{\rm GeV}}{f} \frac{Y_a}{70},
    \label{eq:rhoa}
\end{align}
where $Y_a = q_a/s$.
In addition, the axion abundance can receive contributions sourced by the inhomogeneous axion component generated in a tunneling or phase-mixing process completing the charge transfer between the AD field and the axion field. Since this process occurs only after most of the charge has already been transferred to the axion field, we will focus only on the homogeneous axion component and the production via kinetic misalignment here. This gives a lower bound for the total axion abundance produced.
In the following, we demonstrate how a large enough $Y_{a}$ can be obtained to explain the observed dark matter abundance via kinetic misalignment.

\paragraph{Efficient transfer.} Let us first assume an efficient transfer from the AD field to the axion field. The transfer to the axion field completes when $r_0 \simeq f$ and $q_{a} \simeq m f^2$ at this point. If the entropy production around the completion of the charge transfer is negligible,  $Y_{a}$ is the same as the initial $Y_{\rm \phi}$, which can be easily large enough to produce axion dark matter by kinetic misalignment.

As discussed in Sec.~\ref{subsec:PT}, the charge transfer may be completed by a tunneling process in the thermal potential. From the local minimum at $r_\text{eq} < r_0 \simeq f$ the AD field can tunnel to the true minimum at $r = 0$. 
Let us parameterize the AD field value when the tunneling occurs by $r = X T$. The parameter $X$ ($> 1$ and $\ll f/T$) can be computed once the tunneling rate has been determined.
The energy density of the thermal bath is then bounded from below by $X^2 m^2 T^2$. 
From this, we obtain
\begin{align}
   Y_a \lesssim 6 \times 10^5 \left(\frac{f}{10^9~{\rm GeV}} \frac{100~{\rm TeV}}{m}\right)^{2} \left(\frac{g_*}{200}\right)^{1/2} \left(\frac{10}{X}\right)^{3}
\end{align}
where $g_*$ is the number of effective degrees of freedom of the thermal bath. The bound is saturated when the entropy of the universe is dominantly created by the tunneling process.
Requiring that the maximal possible $Y_{a}$ can explain the axion dark matter by kinetic misalignment, we obtain an upper bound on the mass of the AD field,
\begin{align}
\label{eq:mup_rf}
    m \lesssim 10^4~{\rm TeV} \left(\frac{f}{10^9~{\rm GeV}}\right)^{1/2} \left(\frac{g_*}{200}\right)^{1/4} \left(\frac{10}{X}\right)^{3/2}.
\end{align}
Interpreting the AD field as a supersymmetric flat direction, this upper bound does not require low scale supersymmetry.

\paragraph{Inefficient transfer.} The transfer may not be efficient when $r_0$ reaches $f$. In this case, the transfer completes for $r_0 < f$. The value of such $r_0$ depends on the model, but let us consider a representative case where a process with $\gamma = \epsilon T$ is the bottleneck process. Since the transfer occurs by an interaction coming into equilibrium, the transfer is a non-thermal process and entropy is produced.  Using the following relations at the completion of the transfer,
\begin{align}
    m^2 r_0^2 < \frac{\pi^2 g_*}{30} T^4,~ \epsilon \frac{T^3}{r_0^2} = H(T),  
\end{align}
we obtain
\begin{align}
    Y_{a} \lesssim 1000 \left( \frac{10~{\rm TeV}}{m}\right)^{1/3} \left(\frac{200}{g_*}\right)^{1/2} \left(\frac{\epsilon}{10^{-2}}\right)^{1/3},
\end{align}
where we use as reference value the strong sphaleron process, $\epsilon = 100 \alpha_3 \simeq 10^{-2}$. The upper bound on $m$ from successful DM production from kinetic misalignment is 
\begin{align}
\label{eq:mup_transferr}
    m \lesssim 10^5~{\rm TeV} \left(\frac{10^9~{\rm GeV}}{f}\right)^3 \left(\frac{200}{g_*}\right)^{3/2}\frac{\epsilon}{10^{-2}}.
\end{align}
For $f < 10^{11}$ GeV,
the upper bound
does not require low scale supersymmetry
if the strong sphaleron process is indeed the bottleneck process. The bound can be strong enough to have implications for low scale supersymmetry if a less efficient process is the bottleneck process.
For example, explicit $R$ symmetry breaking provided by the gaugino mass $m_\lambda$ may be required for the transfer to be free-energetically favored. In this case, $\gamma \sim 0.1 m_\lambda^2/T$, and we obtain a bound
\begin{align}
   \frac{m^3}{m_\lambda^2}  \lesssim 40 {\rm TeV} \left(\frac{10^9~{\rm GeV}}{f}\right)^5  \left(\frac{200}{g_*}\right)^{3/2}.
\end{align}
The compatibility with $m,~m_\lambda >$ TeV requires $f < $ few $10^9$ GeV. 
In the context of the MSSM (or its extension), the transfer rate can be computed for different flat directions. The rate may also depend on the sfermion mixing, since this can enhance chiral symmetry breaking rate~\cite{Co:2021qgl}. We leave the detailed estimation of the transfer rate for future work.

\subsection{Baryogenesis}

A rotating axion field generically induces a non-zero baryon number as long as the baryon number violation from the electroweak sphaleron process is efficient~\cite{Co:2019wyp,Domcke:2020kcp,Co:2020xlh}.
The baryon number is frozen after the electroweak phase transition.

For the QCD axion, however, the axion velocity that can explain the observed baryon asymmetry produces too much axion dark matter via the kinetic misalignment mechanism~\cite{Co:2019wyp}. This problem can be avoided by introducing additional BSM physics which raises temperature of the electroweak phase transition in comparison with the standard model prediction.

Alternatively, the mass of an axion-like particles can be lighter than that of the QCD axion, so that the overproduction of dark matter is avoided, see Eq.~\eqref{eq:rhoa}. However, in this case it turns out to be non-trivial to transfer the charge of the AD field to the axion-like particle. The transfer requires the violation of the shift symmetry of the axion-like particle, see Sec.~\ref{sec:zero}. If this is induced by the QCD anomaly, this also gives a mass to the axion-like particle and dark matter is overproduced by kinetic misalignment. One can additionally introduce the QCD axion which couples to QCD more strongly than the axion-like particle does, so that the axion-like particle does not obtain a mass. However, when the QCD anomaly induces the rotation of the axion-like particle, the QCD axion also rotates, and the overproduction occurs.
If the violation of the shift symmetry is instead induced by the weak anomaly, the axion-like particle does not obtain too large a mass. However, within the standard model, electroweak sphalerons are the only interaction which violates $B+L$ through the weak anomaly,  so the transfer into the axion-like particle requires large asymmetry of $B+L$ fermions and thus does not minimize the free energy. This can be avoided if the initial rotation of the AD field carries $B+L$ charge and can thus absorb the change in the total $B+L$ charge without invoking large fermion asymmetries or if there are additional $B+L$ violating interactions.
In the former case, the $B-L$ charge of the AD field should be zero to avoid the overproduction of baryon asymmetry.
In the latter case, 
the extra $B+L$ violation can directly create a baryon asymmetry if $B-L$ is simultaneously violated~\cite{Domcke:2020kcp,Co:2020jtv} (see also~\cite{Co:2021qgl,Harigaya:2021txz,Chakraborty:2021fkp,Kawamura:2021xpu}), so the production from the axion motion through the electroweak sphaleron process may be subdominant. We conclude that if the axion-like particle couples to the weak anomaly,
there is explicit
$B+L$ violation in the system (which initiates the rotation of the AD field or washes-out $B+L$ charges)
and $B-L$ is (approximately) preserved,
then the motion of this axion-like particle at the electroweak phase transition could be responsible for the observed matter-antimatter asymmetry of the universe without overproducing dark matter.

\subsection{Kination domination by axion rotation}

Since the energy density of the rotation of the AD field decreases as $R^{-3}$, if the initial field value of the AD field is sufficiently large, the rotation energy of the AD field can come to dominate over the energy of the thermal bath. We show that in this case the universe can enter a kination-dominated era driven by the axion rotation.

\paragraph{From an AD field era to a kination era.} Let us start from the initial state where the universe is dominated by the AD field and almost all of the charge $q > q_c$ is in the AD field, with the initial charge in the axion much below the equilibrium value $\sim m f^2$.
Once the charge transfer becomes efficient and the axion rotation reaches equilibrium, the total energy density of the AD and axion fields decreases by $m^2 f^2/2$. This amount of energy should go to the thermal bath and entropy is created. If the thermal bath has an energy density smaller than this before this entropy production, the temperature of the universe when the axion and AD field begin to be in chemical equilibrium is determined by $T^4\sim m^2 f^2$. Otherwise, the temperature of the bath remains approximately constant up to cosmic expansion.

As long as the charge transfer between the axion and AD fields continues to be efficient, the system evolves adiabatically. In particular, the axion and AD fields follow the equilibrium values $\dot{\theta_a} = -m$ and $r^2 = r_0^2 - f^2$, and the entropy and energy density of the thermal bath decrease as $R^{-3}$ and $R^{-4}$, respectively. Thus, the universe continues to be dominated by the AD field.

The energy density of the axion and AD fields is given by
\begin{align}
    \rho = \rho_a + \rho_\phi =\frac{1}{2}\dot{\theta}_a^2 f^2 + m^2 r^2 = m q - \frac{1}{2}m^2 f^2,
\end{align}
and decreases following
\begin{align}
    \frac{{\rm dln} \rho}{{\rm dln} R} = \frac{-3  q}{q - m f^2/2} \rightarrow
    \begin{cases}
    -3 & : q \gg q_c = m f^2 \\
    -6  & :q \rightarrow q_c 
    \end{cases},
\end{align}
where we used the charge conservation $q\propto R^{-3}$. For $q \gg q_c$, the equation of state is that of matter, while as $q$ approaches $q_c$, it approaches that of kination. 
The charge transfer completes at $q = q_c$, after which the universe is dominated by the axion rotation with $\dot{\theta}_a \propto R^{-3}$, and the universe is kination-dominated.

\paragraph{Implications.} A kination-dominated era due to a rotating axion, called axion kination, is realized in~\cite{Co:2019wyp} starting from the rotation of a PQ symmetry breaking field that has a nearly quadratic potential. Our realization is applicable to more generic potentials of the PQ symmetry breaking field since the rotation is initiated in another sector.
The duration of the kination-dominated era depends on the amount of the radiation energy density at the completion of the charge transfer, and thus on the preceding cosmological history, e.g., the initial field value of the AD field, the reheating temperature, and when the AD field is thermalized. Also, as we discussed in Sec.~\ref{sec:tunneling}, the very end of the charge transfer can involve tunneling process that creates small amount of entropy, which will also limit the duration of the kination-dominated era. We leave an investigation of the duration of the kination era to future work.

In the above discussion, we assume $ m \ll T$ to argue that the energy density of the bath decreases as $R^{-4}$. If $m$ is larger than $T$, the chemical potential $\mu \sim m > T$ ensures that the energy density remains constant $\sim m^4$. This is smaller than the axion energy density, $m^2 f^2$, as long as $m \ll f$ and hence the axion kination era still occurs. Note that in this limit, scalar fields do not receive thermal masses larger than $\mu \sim m$. If there exist an additional scalar field that can receive charges from the AD field and has a smaller energy per charge, the scalar field necessarily is destabilized by the negative mass from the chemical potential; see also Sec.~\ref{sec:conclusion}.

\subsection{Gravitational waves}

The charge transfer between an AD field and an axion field offers two interesting possibilities for gravitational wave observations. First, any gravitational waves produced prior to a kination era (such as those produced by inflation or cosmic strings) are enhanced~\cite{Giovannini:1998bp,Cui:2017ufi}. As in the setup in~\cite{Co:2019wyp}, the kination-dominated era in our setup is preceded by a matter-dominated era. This leaves peculiar signatures in the spectrum of primordial gravitational waves~\cite{Co:2021lkc,Gouttenoire:2021wzu,Gouttenoire:2021jhk}. Second, if the charge transfer occurs via a first order phase transition (see discussion in Sec.~\ref{subsec:PT}), the associated bubble dynamics source gravitational waves (see~\cite{Caprini:2019egz} for a recent review.)
The magnitude of the gravitational waves will depend on the detail of the phase transition, such as the latent heat and the duration of the phase transition, motivating the detailed investigation of the dynamics.

\section{Summary and Discussion}
\label{sec:conclusion}

In this paper, we discussed charge transfer between complex scalar fields. A rotating complex field $\phi$ has a non-zero $U(1)_\phi$ charge. This $U(1)$ charge can be transferred into a $U(1)_P$ charge of another complex scalar field $P$ so that $P$ begins rotation if $U(1)_\phi\times U(1)_P$ is explicitly broken 
with only one linear combination of the original charges conserved.
We considered the charge transfer through couplings to a thermal bath and focused on the case where $\phi$ has a nearly quadratic potential with a mass $m$, the angular direction of $P$ is an axion field, and the radial direction of $P$ is strongly fixed at a constant value $f$. Whether or not the $U(1)$ charge is dominantly in $\phi$ or $P$ is determined by minimizing the free energy. We found that almost all of the $U(1)$ charge is transferred into the axion field once the total charge has been redshifted to a critical value of about $m f^2$ as long as no conservation law requires the state with large $U(1)$ charge in $P$ to have a large fermion asymmetry. The latter condition requires that all linear combinations of $U(1)_\phi$ and chiral symmetry as well as those of $U(1)_P$ and chiral symmetry be explicitly broken. We moreover find that the transfer rate is suppressed by the ratio between the temperature and the smaller of the radii of $\phi$ and $P$ as well as by explicit symmetry breaking rates.

The scenario has an immediate phenomenological application. The kinetic energy of the induced axion rotation, if large enough, contributes to the axion dark matter density through so-called the kinetic misalignment mechanism. Unlike the original realization in~\cite{Co:2019jts}, our case does not require a flat potential of the radial direction of the PQ symmetry breaking field, and thus is compatible with a wider class of PQ symmetry breaking models including dynamical symmetry breaking models~\cite{Choi:1985cb}.
\footnote{In supersymmetric models~\cite{Harigaya:2015soa}, the PQ charged scalar degree of freedom may have a flat potential at field values much above the dynamical scale. However, the potential may be flatter than a quadratic one and the dynamics of the PQ field be complicated by Q-ball formation.}
This comes with the advantage that in dynamical PQ symmetry breaking models, the smallness of the PQ breaking scale compared to the Planck scale is understood by dimensional transmutation and the PQ symmetry is more easily understood as an accidental symmetry than the models with a fundamental PQ scalar~\cite{Randall:1992ut}. 
Also, the axion rotation can be induced even when the PQ breaking occurs after cosmic inflation, leading to cosmic string configurations rotating in field space. See~\cite{Co:2022qpr} for the existence of rotating cosmic string solutions. This may impact the axion dark matter abundance produced from the string-domain wall network.

Further phenomenological implications are tied to the detailed dynamics of the charge transfer. 
Contributions to the effective potential of $\phi$ induced by particles coupling to $\phi$ favour $\phi =0$, and hence the state with most charge in $P$.
Once the total charge density becomes sufficiently small through cosmic expansion, the state with most charge in $\phi$ becomes a metastable state and the state with most charge in $P$ becomes the absolute minimum. The charge transfer can then involve a first order phase transition or phase mixing. The former will source gravitational waves. Both will create fluctuations of the axion field, which can contribute to axion dark matter. We hope that these observations can serve as a motivation for a more detailed investigation of the dynamics of this charge transfer.

Our work can be immediately generalized to charge transfer process between generic complex fields. Transfer is possible as long as it is favored by the free energy and the transfer rate exceeds the Hubble expansion rate. For example, charge transfer can occur from a MSSM flat direction to another flat direction, if no conservation law requires a large fermion asymmetry and if the latter flat direction has a smaller energy per charge. It is also possible that the final state is composed of rotations of several flat directions. The charge transfer may complete by first order phase transition or phase mixing, depending on the thermal potentials of the fields involved in the charge transfer.
We note that the axion case, i.e., a recipient field for the charge with a fixed value for the radial component, is special in the sense that the energy per charge of the axion rotation $\sim q/f^2$ is always smaller than that of a complex scalar in a quadratic potential for a sufficiently small total $U(1)$ charge.

\section*{Acknowledgement}
K.M.~was supported in part by MEXT Leading Initiative for Excellent Young Researchers Grant No.~JPMXS0320200430.

\appendix

\section{Boltzmann equations for charge transfer}
\label{app:Boltzmann}

\begin{figure}[t]
\centering
 \includegraphics[width = 0.45 \textwidth]{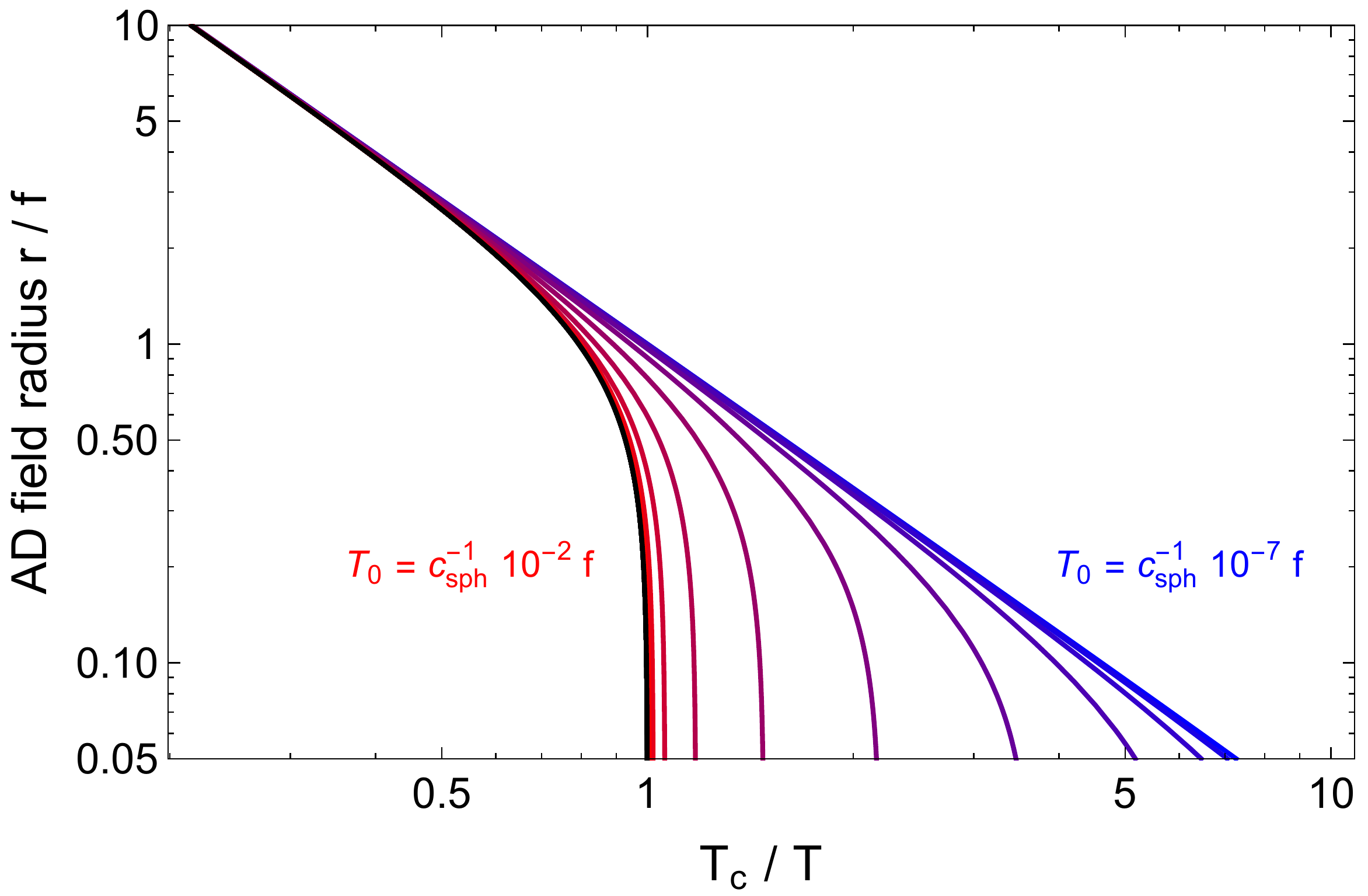} \hfill
  \includegraphics[width = 0.45 \textwidth]{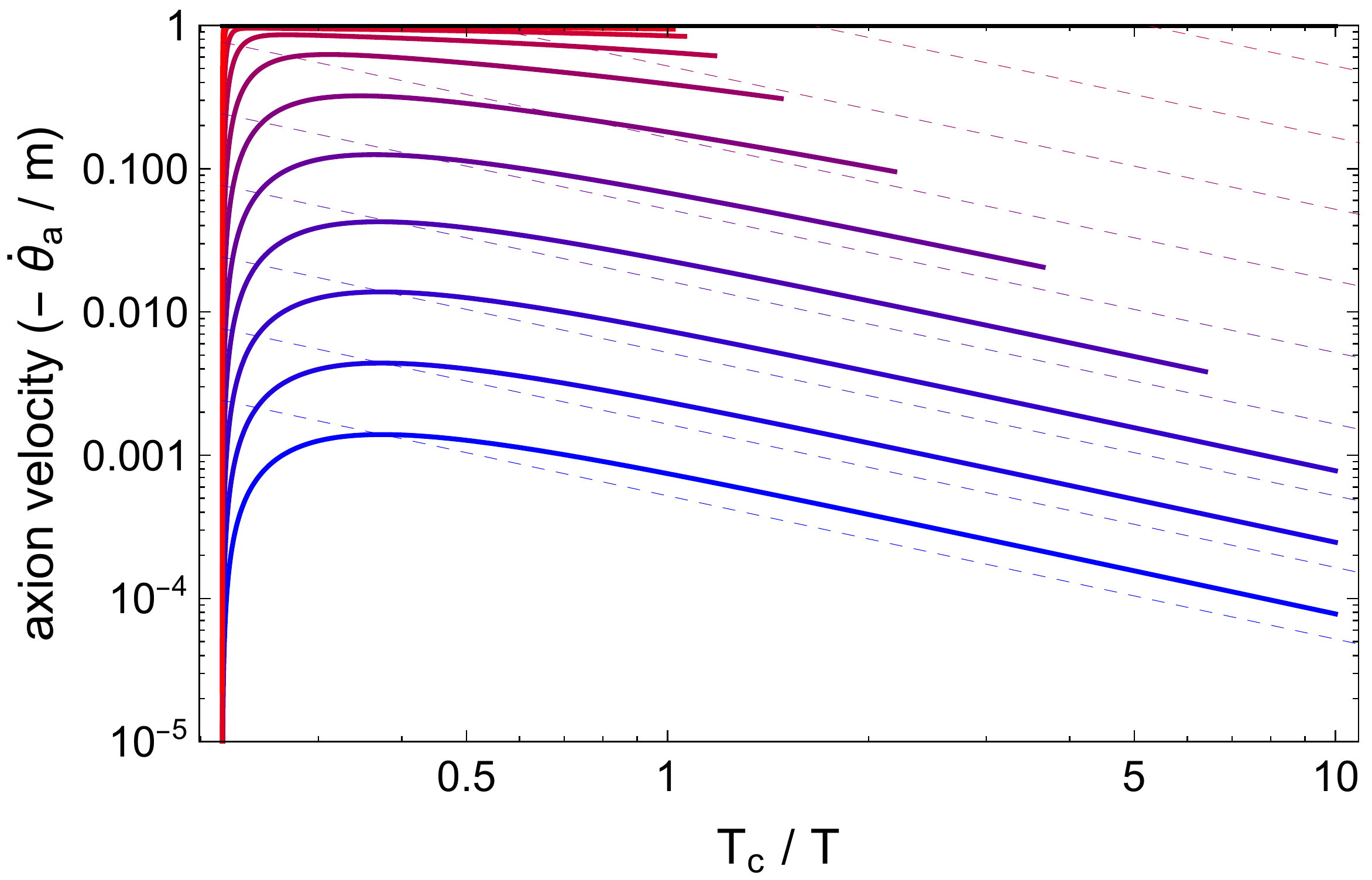} 
 \caption{Evolution of charges stored in the AD field and the axion, governed by the AD field radius $r$ and the axion velocity $\dot \theta_a$ for different initial temperatures $T_0$. The black curve depicts their equilibrium values, obtained if the charge transfer rate is significantly faster than the Hubble expansion. Here cosmic expansion is accounted for but thermal contributions to the effective potential are not included.}
 \label{fig:Boltzmann}
\end{figure}

In this appendix we show numerical results obtained by solving the Boltzmann equations~\eqref{eq:Boltzmann1} for the model discussed in Sec.~\ref{sec:gauge}. We fix $f = 10^9$~GeV, $m = 10^3$~GeV, $\alpha_G = 0.1$ and consider a radiation-dominated background, $H = T^2/M_*$ with $M_* = 7 \cdot 10^{17}$~GeV. We initialize the AD field at $r = r_0 = 10 f$, treating the temperature $T_0$ at this time as a free parameter. We denote the temperature at which the AD field would collapse assuming an efficient charge transfer ($q = q_c$) as $T_\text{c}$. 
The results are shown in Fig.~\ref{fig:Boltzmann} as a function of the temperature of the thermal bath $T$.
 
The left panel of Fig.~\ref{fig:Boltzmann} shows the evolution of the radial component of the AD field (recall that the angular component is fixed to $\theta_\phi = m$). For low temperatures, the charge transfer is not efficient, and we observe that the field value decreasing as $r  \simeq  r_0 \propto R^{-3/2}$, in agreement with the conserved charge $R^3 r_0^2 m$ in the absence of charge transfer. At larger temperatures the charge transfer becomes efficient, tracing the equilibrium value for $r$ for $\gamma_s (T/f)^2 > H$. Once we reach $r \sim f$ the AD field collapses if the charge transfer is efficient, otherwise the collapse is delayed until the Hubble rate falls below the charge transfer rate.

The right panel shows the evolution of the axion velocity, starting from $\dot \theta_a = 0$. The equilibrium value $\dot \theta_a = -m$ is reached if the transfer is efficient at $T = T_0$. More generally the axion velocity is well described by Eq.~\eqref{eq:theta_a_inefficient_transfer}, as indicated by the dashed lines. The curves for larger values of $T_0$ end once the AD field collapses, and all charge is transferred to the axion, see left panel. After this occurs, $\dot{\theta}_a$ should decrease in proportion to $R^{-3}$.

In the model discussed in Sec.~\ref{sec:fermion}, the charge density for the fermions efficiently tracks the equilibrium value given in Eq.~\eqref{eq:boltzmann_mass-eq} as long as at least one of the two interactions rates $\gamma_s$ and $\gamma_\chi$ is faster than the Hubble expansion. In this case, after replacing the $q_\chi$ by its equilibrium value in Eq.~\eqref{eq:boltzmann_mass}, the dynamics of the axion and AD field are analogous to the the model of Sec.~\ref{sec:gauge}. For the impact of thermal corrections, see Sec.~\ref{sec:tunneling}.

\section{Thermodynamics of rotating scalar fields}
\label{app:Veff}

The main purpose of this section is to demonstrate how to obtain thermodynamic quantities in the presence of rotating scalar fields by performing explicit calculations in a simplified model. Unlike the simplified setup in Sec.~\ref{sec:tunneling}, we include an AD-field dependent mass term for the fermion which participates in the charge transfer between the AD and the axion field. We moreover include the fluctuations of the AD field. Nevertheless, we find a qualitatively similar result as in Sec.~\ref{sec:tunneling}.

\subsection{Preliminary}

\paragraph{Model.}
To make our discussion concrete, we consider the following toy model,
\begin{align}
  \mathcal{L}
  =& \abs{\partial \phi}^2 - m^2 \abs{\phi}^2 + \overline{\psi} i \slashed{D} \psi
  + \frac{1}{2} f^2 \qty( \partial \theta_a )^2 - \frac{1}{2} \Tr G_{\mu\nu} G^{\mu\nu} \nonumber \\
  &- \qty( y \phi \overline{\psi} P_L \psi + \text{H.c.} ) - \frac{g^2 \theta_a}{16 \pi^2} \Tr G_{\mu\nu} \tilde G^{\mu\nu}.
  \label{eq:toy_app}
\end{align}
Here $\phi$ is a complex scalar field whose charge is transferred to the axion $\theta_a$ in the end,
and $\psi$ is a Dirac fermion charged under a gauge group with field strength tensor $G_{\mu \nu}$. The chiral symmetry of the fermion is explicitly broken by the chiral anomaly. Through the Yukawa coupling $y$, the fermion obtains a $\phi$-dependent mass. In a supersymmetric theory, where a nearly quadratic potential of $\phi$ can be achieved, $y$ is as large as the gauge coupling constant, so we assume  $y = {\cal O}(1)$ in the following.
The relevant current equations are
\begin{align}
  \partial \cdot J_\phi + \frac{1}{2} \partial \cdot J_5 = - \frac{g^2}{16 \pi^2} \Tr G_{\mu\nu} \tilde G^{\mu\nu}, \qquad
  \partial \cdot J_a = - \frac{g^2}{16 \pi^2} \Tr G_{\mu\nu} \tilde G^{\mu\nu},
\end{align}
where each current is defined by
\begin{align}
  J_\phi^\mu = \phi^\dag i \overleftrightarrow{\partial} \phi, \qquad
  J_5^\mu = \overline{\psi} \gamma^\mu \gamma_5 \psi, \qquad
  J_a^\mu = f^2 \partial^\mu \theta_a.
\end{align}
This Lagrangian has a U$(1)$ symmetry, whose conserved charge is given by
\begin{align}
  Q = Q_\phi + \frac{1}{2} Q_5 - Q_a,
\end{align}
with $Q_\bullet = \int_{\bm{x}} J_\bullet^0$.
We also define charge densities for later convenience as $q_\bullet = \langle Q_\bullet \rangle / \mathbb{V}$ with $\mathbb{V}$ being a spatial volume.
On top of Eq.~\eqref{eq:toy_app}, we introduce a higher dimensional operator for $\phi$ that explicitly breaks this U$(1)$ symmetry.
When the field value of $\phi$ is large in the early Universe, this breaking term induces the rotation of $\phi$ via the Affleck-Dine mechanism. 
After some time, the breaking becomes inefficient as the field value of $\phi$ decreases due to the cosmic expansion, and then $Q$ becomes conserved.

When the field value of $\phi$ is large enough, one may integrate out the Dirac fermion $\psi$.
The Lagrangian in this case reads
\begin{align}
  \mathcal{L}
  =& \frac{1}{2} \qty(\partial r)^2 - \frac{1}{2} \qty( m^2 - \qty(\partial \theta_\phi)^2 ) r^2 + \frac{1}{2} f^2 \qty( \partial \theta_a )^2 - \frac{1}{2} \Tr G_{\mu\nu} G^{\mu\nu} 
  - \frac{g^2 \qty(\theta_a + \theta_\phi)}{16 \pi^2} \Tr G_{\mu\nu} \tilde G^{\mu\nu},
\end{align}
and the current equations are
\begin{align}
  \partial \cdot J_\phi = - \frac{g^2}{16 \pi^2} \Tr G_{\mu\nu} \tilde G^{\mu\nu}, \qquad
  \partial \cdot J_a = - \frac{g^2}{16 \pi^2} \Tr G_{\mu\nu} \tilde G^{\mu\nu}.
\end{align}
We can see that the model \eqref{eq:toy_app} is a simple realization of the toy model discussed in Sec.~\ref{sec:gauge}.

\paragraph{Review of thermal field theory.}
Before going into our model calculations, we briefly summarize some basic facts of thermal field theory~\cite{Kapusta:2006pm,Laine:2016hma}.
Starting from the Hamiltonian $\Ham$, thermodynamic quantities with a conserved charge $Q$ are obtained from the grand canonical ensemble of 
\begin{align}
  \rho_\text{GC} = \frac{1}{Z} e^{- \frac{1}{T} \qty( \Ham - \mu Q )},
  \qquad Z = \Tr \rho_\text{GC},
\end{align}
which is related to the thermodynamic pressure 
\begin{align}
  p (T,\mu) = \frac{T}{\mathbb{V}} \ln Z,
\end{align}
with $\mathbb{V}$ being a spatial volume.
The charge density is obtained from \footnote{
In the main text, we often simply refer to $q$ as `charge' for brevity.
}
\begin{align}
  q = \frac{\partial p \qty(T,\mu)}{\partial \mu}.
\end{align}
These quantities are useful when considering a bath of fixed temperature $T$ and chemical potential $\mu$.
However, we are rather interested in the case of a fixed charge $\langle Q \rangle$.
The Helmholtz free energy, $f (T, q)$, is more useful in this case, which is obtained from the Legendre transformation
\begin{align}
  f \qty(T, q) = \mu q - p.
\end{align}

In the context of field theory, it is more convenient to express these quantities by means of an effective potential, \textit{i.e.}, as a function of $\phi$.
Thermodynamic quantities are obtained by evaluating it at its extrema,
\begin{align}
  p \qty(T, \mu) = - \left. V_{\mu, \text{eff}} \qty(\phi, T ,\mu) \right|_{\dd V_{\mu, \text{eff}} / \dd \phi = 0},
  \qquad
  f \qty(T, q) = \left. V_\text{eff} \qty(\phi,T, q) \right|_{\dd V_\text{eff} / \dd \phi = 0}.
\end{align}
with the subscript indicating the thermodynamic quantity taken to be constant.
These effective potentials are related through the Legendre transform of
\begin{align}
  V_\text{eff} \qty(\phi, T, q) = \mu q + V_{\mu, \text{eff}} \qty(\phi, T, \mu),
  \qquad
  q = - \frac{\partial V_{\mu, \text{eff}} \qty(\phi,T,\mu)}{\partial \mu},
\end{align}
Note that we assume that the background $\phi$ is homogeneous.
The grand canonical partition function can be expressed as an Euclidean path integral
\begin{align}
  Z = e^{\frac{\mathbb{V}}{T} p \qty(T, \mu) }
  = \int_\text{b.c.} \qty[\dd \mu]\, 
  \exp \qty[ - \int_0^{\frac{1}{T}} \dd \tau \dd^3 x\,  \mathcal{L}_\text{E} \qty( \partial_\tau \to \partial_\tau - c_\phi \mu )
  ],
\end{align}
with $c_\phi$ denoting the charge of $\phi$ with respect to the conserved charge $Q$.
The boundary condition (b.c.) is taken so that the bosonic fields are periodic while fermionic fields are anti-periodic.
The effective potential $V_\mu$ is obtained by performing this path integral on a background of $\phi$.
Then we obtain the second effective potential $V_q$ via the Legendre transform.

\subsection{Effective potential}

\paragraph{Without axion.}
We first consider the case where the conversion of charge between $\phi$ and $\theta_a$ mediated by the sphaleron is negligible.
The axion is sequestered and hence we omit it hereafter, as in Sec.~\ref{sec:f w/o axion}.
The relevant conserved charge is then
\begin{align}
  Q_\text{AD} = Q_\phi + \frac{Q_5}{2}.
\end{align}
The Euclidean Lagrangian reads
\begin{align}
  \mathcal{L}_\text{E} = 
  ( \partial_\tau \phi^\dag + \mu \phi^\dag)  ( \partial_\tau \phi - \mu \phi ) + \abs{\nabla \phi}^2 + m^2 \abs{\phi}^2
  + \overline{\psi} \qty( \gamma^0 \partial_\tau - i \gamma^i \partial_i + \frac{\mu}{2}\gamma^0 \gamma_5 ) \psi + 
  \qty( y \phi\overline{\psi} P_L \psi + \text{H.c.} ).
\end{align}
Assuming a homogeneous background of $\phi = r e^{i\theta_\phi} / \sqrt{2}$, one may compute its effective potential
\begin{align}
  V_{\mu, \text{eff}} \qty(r, T, \mu) = \qty( m^2 - \mu^2 ) \frac{r^2}{2} - d_\psi \ln \det \qty( i \slashed{\partial} + \frac{yr}{\sqrt{2}} + \frac{\mu}{2} \gamma^0 \gamma_5 ) 
  +
  \ln \det \qty( \Box + m^2 - \mu^2 + \mu i \overleftrightarrow{\partial_t}  )
  \label{eq:effmu-AD}
\end{align}
at the one-loop level. Here $d_\psi$ counts the number of Dirac fermions (e.g., $d_\psi = 1/2$ for a Weyl fermion, $3$ for a Dirac color triplet, ...). The determinants are evaluated as
\begin{align}
    \ln \det \qty( i \slashed{\partial} + \frac{yr}{\sqrt{2}} + \frac{\mu}{2} \gamma^0 \gamma_5 )  = & 2\int \frac{\dd^3 p}{(2\pi)^3} \left[ \frac{\omega'}{2} + T \ln \left(1 +  e^{-\omega'/T}\right)
    \right]_{\omega' = \sqrt{y^2 r^2/2 + (p-\mu/2)^2}} + \left(\mu \rightarrow - \mu \right),   \nonumber \\
    = & \int \frac{\dd^3 p}{(2\pi)^3} \left[ \frac{\omega'}{2} + T \ln \left( 1 -  e^{-\omega'/T} \right) \right]_{\omega' = \sqrt{m^2 + p^2} - \mu} + \left(\mu \rightarrow - \mu \right).
\label{eq:int_f}
\end{align}
for the fermionic and bosonic contribution, respectively.
The first terms in the integrands lead to UV divergence, which are the usual zero temperature divergent parts and can be dropped.%
\footnote{The divergence in the fermion contribution includes the logarithmic divergence of the coefficient of $\mu^2 r^2$. One may worry about this divergence at finite densities, but this contribution is cancelled by the wave-function renormalization of $\phi$, owing to the non-renormalization of the charge.}
From~\eqref{eq:int_f},  the approximate form of the fermion contribution is
\begin{align}
    V_{\mu,f} / d_\psi \simeq
    \begin{cases}
     - 4 \frac{7}{8} \frac{\pi^2 T^4}{90}
  - \frac{\mu^2  T^2}{24} - \frac{\mu^4}{192 \pi^2}  + \frac{y^2 T^2 r^2}{24}   & yr \ll T, \\
  -\frac{\sqrt{2}}{\pi^{3/2}} e^{-\frac{y r}{\sqrt{2} T}} \left(\frac{y r}{\sqrt{2}}\right)^{3/2} T^{5/2} -
  \frac{3}{4 \sqrt{2} \pi^2}e^{-\frac{y r}{\sqrt{2} T}} \sqrt{ \frac{y r}{\sqrt{2}} T^3} \mu^2 -\frac{\mu^4}{192 \pi^2}
        & yr \gg T.
    \end{cases}
\end{align}
For the fluctuations of the AD field, \eqref{eq:int_f} yields
\begin{align}
    V_{\mu,{\rm AD}} \simeq - \frac{\pi^2}{45} T^4 - \frac{1}{6} \mu^2 T^2,
\end{align}
where we have assumed $m, \mu < T$ and have dropped  higher-order terms in $\mu$.

Around $r = 0$, assuming $\mu <T$ (dropping $O(\mu^4)$ terms), 
\begin{align}
    V_{\mu, \text{eff}} (r, T, \mu) \simeq \qty( m^2 - \mu^2 ) \frac{r^2}{2} 
  - d_\psi \qty( 4 \frac{7}{8} \frac{\pi^2 T^4}{90}
  + \frac{\mu^2  T^2}{24} + \frac{\mu^4}{192 \pi^2}  - \frac{y^2 T^2 r^2}{24} )  
  - \qty( \frac{\pi^2 T^4}{45} + \frac{\mu^2 T^2}{6}).
\end{align}
Using the Legendre transform, the effective potential for a fixed charge density $q$ is
\begin{align}
    V_\text{eff} (r,T,q) = - \frac{\pi^2}{45} \qty(1 + \frac{7}{4}d_\psi) T^4 + \frac{1}{2} \qty(m^2 + \frac{d_\psi}{12}y^2 T^2) r^2 + \frac{6 q^2}{12 r^2 + (4 + d_\psi) T^2}.
\end{align}
This has an extremum at $r=0$. The second derivative of $V_q$ with respect to $r$ at $r=0$ is
\begin{align}
   \left( \frac{ \partial ^2V_\text{eff} (r, T, q)}{\partial r^2}\right)_{r=0} = m^2 + \frac{d_\psi}{12} y^2 T^2 - \frac{144 q^2}{(4+ d_\psi)^2 T^4}.
\end{align}
This is positive and the origin is a (local) minimum if the charge $q$ is small enough,
\begin{align}
    q^2 < \frac{(4+ d_\psi)^2 d_\psi}{1728} y^2T^6 + \frac{(4+ d_\psi)^2 }{144} m^2T^4
\end{align}
This condition is qualitatively the same as the one derived below Eq.~(\ref{eq:Veff_comparison}). Note that $q$ satisfying this condition implies $\mu < T$, and hence our assumption of $\mu < T$ is self-consistent. When $q$ significantly violates this condition, $\mu^2 > m^2 + d_\psi y^2 T^2/12$. The integration in Eq.~(\ref{eq:int_f}), even after the resummation to include the thermal mass $\sim y T$, results in an imaginary part, signaling the instability of $r=0$ and leading to the formation of the BEC of the AD field.

For $y r \gg T$, assuming $\mu < T$,
\begin{align}
    V_{\mu, \text{eff}}(r, T, \mu) \simeq \qty( m^2 - \mu^2 ) \frac{r^2}{2} - \qty( \frac{\pi^2 T^4}{45} + \frac{\mu^2 T^2}{6}).
\end{align}
The effective potential for a fixed charge density $q$ is
\begin{align}
    V_\text{eff} (r,T,q) = - \frac{\pi^2}{45} T^4 + \frac{1}{2}m^2  r^2 + \frac{3 q^2}{6 r^2 + 2 T^2}.
\end{align}
This has an extremum at $r^2 = |q|/m \equiv r_{\rm eq}^2$. The second derivative at $r_{\rm eq}$ is $4m^2$. The assumption of $y r_{\rm eq} \gg T $ is valid if $q \gg m T^2/ y^2 $. At $r_{\rm eq}$, $|\mu| = m$, so the integration in Eq.~(\ref{eq:int_f}) does not yield an imaginary part and $V_q$ can be interpreted as the effective potential as usual.  Moreover, for $m <T$, the assumption $\mu = m < T$ is again self-consistent. For $m > T$, Eq.~\eqref{eq:effmu-AD} is simply dominated by the first term, and thermal corrections are irrelevant to the dynamics of the phase transition.

There are two minima of the effective potential for $ m T^2/ y^2 < q <  y T^3$. As in Sec.~\ref{sec:f w/o axion}, one can compare the effective potential at these two minima and determine which is the absolute minimum, obtaining qualitatively the same result.

We are interested in the possible transition from $r= r_{\rm eq}$ to $r=0$ by bubble nucleation. To compute the nucleation rate, one needs the effective potential for intermediate field values, $0 < r < r_{\rm eq}$. When $|q| > m T^2$ and $r < r_{\rm eq}$, $|\mu|$ is larger than $ m$. Also, for $y r  > T$, the thermal mass squared of the AD field given by the coupling $y$ is negative. Therefore, for $T/y<r < r_{\rm eq}$, the integration in Eq.(\ref{eq:int_f}) yields an imaginary part, and the effective potential inside the bubble wall formally contains an imaginary part. This may cast a doubt on the validity of the computation, but we argue that the appearance of the imaginary part is an artifact of the inclusion of long-wave length modes despite the finiteness of the width of the bubble wall~\cite{Boyd:1992xn}. In computing the effective potential, the field value $r$ is approximated to be homogeneous, which is justified only for the excitations with wavelength shorter than the wall width. The effective potential contains a term $- \mu^2(q,r) r^2/2$, so the width of the bubble-wall is at most $O(\mu^{-1})$. We should, therefore, use a prescription where $p < \mu$ is excluded from the integration in Eq.~(\ref{eq:int_f}). This regularization prescription introduces some uncertainty in the coefficient of $\mu^2 T^2$ terms; they depend on the choice the IR cut off. However, since we are interested in $r > T/y > T$, for which the $\mu^2 r^2$ term dominates over $\mu^2 T^2$ terms, this IR dependence does not introduce any relevant uncertainty in the computation of the bubble nucleation rate.  An explicit computation of this nucleation rate is beyond the scope of this paper.

\paragraph{With axion.}
Finally, we turn on the sphaleron processes to consider the charge transfer to the axion through a Chern-Simons coupling, as in Sec.~\ref{sec:f w axion}.
In this case, the axion and AD charges are no longer separately conserved. Rather their linear combination is the only conserved quantity of the system, i.e.,
\begin{align}
Q = Q_\text{AD} - Q_a = Q_\phi + \frac{Q_5}{2} - Q_a.
\end{align}
Assuming a homogeneous background of $\phi = r e^{i \theta_\phi} / \sqrt{2}$ again, the one-loop effective potential acquires an additional contribution from the axion
\begin{align}
  V_{\mu, \text{eff}} (r, T, \mu)
  \simeq \left. V_{\mu, \text{eff}} (r, T, \mu) \right|_{\text{Eq.~\eqref{eq:effmu-AD}}}
  - \frac{\mu^2 f^2}{2} + \frac{1}{2} \ln \det \Box .
  \label{eq:effmu-withaxion}
\end{align}
Here we neglect
the potential from the coupling with $G\tilde{G}$ and non-perturbative gauge dynamics,
which makes a particular linear combination of $\theta_\phi - \theta_a$ massive.
The last term comes from the axion fluctuations, which does not lead to a $\mu$-dependent effective potential and hence one may omit it for our purpose.

Using Eq.~(\ref{eq:effmu-withaxion}), we may compute $V_q$ and check the minimum of the effective potential. The result is identical to the discussion in Sec.~\ref{sec:f w axion} up to the modification of $O(1)$ factors.

\small
\bibliographystyle{utphys}
\bibliography{refs}
  
\end{document}